\documentclass[10pt,letterpaper]{elsarticle}

\usepackage{amsmath}
\usepackage{comment}
\usepackage{array,multirow}
\usepackage{float}
\usepackage{bm}
\usepackage{hyperref}
\usepackage{algorithm,algpseudocode}
\usepackage{rotating}

\newcommand{\ffrac}[2]{\ensuremath{\frac{\displaystyle #1}{\displaystyle #2}}}
\newcommand{\p}{\partial}
\usepackage{ulem} 
\usepackage{xcolor} %for color 

\usepackage{soul} 
\setstcolor{red}

\begin{document}

\title{Practical parameter identifiability of respiratory mechanics in the extremely preterm infant}

\author{Richard Foster$^{1}$, Laura Ellwein Fix$^{1}$}

%\address{$^{1}$Department of Mathematics and Applied Mathematics, Virginia Commonwealth University, Richmond, VA, USA}

%\subject{Biomathematics, Mathematical modeling}
%\keywords{Nonlinear dynamics, sensitivity analysis, compartmental modeling, optimization, mathematical physiology, neonatology}
%\corres{Laura Ellwein Fix\\ 
%\email{lellwein@vcu.edu}}

%\keywords{Nonlinear dynamics \and Sensitivity analysis \and Compartmental modeling \and Optimization \and Neonatology \and Mathematical physiology}
%
 %\subclass{92C50 \and 37M05 \and 90C31}

\begin{abstract}
The complexity of mathematical models describing respiratory mechanics has grown in recent years, however, parameter identifiability of such models has only been studied in the last decade in the context of observable data. This study investigates parameter identifiability of a nonlinear respiratory mechanics model tuned to the physiology of an extremely preterm infant, using global Morris screening, local deterministic sensitivity analysis, and singular value decomposition-based subset selection. The model predicts airflow and dynamic pulmonary volumes and pressures under varying levels of continuous positive airway pressure, and a range of parameters characterizing both surfactant-treated and surfactant-deficient lung. Sensitivity analyses indicated eleven parameters influence model outputs over the range of continuous positive airway pressure and lung health scenarios. The model was adapted to data from a spontaneously breathing 1 kg infant using gradient-based optimization to estimate the parameter subset characterizing the patient's state of health. 

\end{abstract}

\maketitle

%% SECTIONS:

% 1: INTRODUCTION
% 2: MODEL
    % (a): State Equations
    % (b): Constitutive Relations
        % (i) Lung scenarios
        % (ii) Airway opening pressure
% 3: DATA (ABBASI)
% 4: MODEL SENSITIVITY AND IDENTIFIABILITY
    % (a): Morris Screening
    % (b): Local Analysis
    % (c): Subset selection
    % (d): Optimization
    % (e) Summary
% 5: RESULTS
    % (a): Morris Screening
    % (b): Model Reduction
    % (c): Local analysis
    % (d): Identifiability
    % (e): Parameter Estimation
% 6: DISCUSSION
% 7: CONCLUSION

%%%%%%%%%% INTRODUCTION

\section{Introduction}\label{sec:intro}

% Current models in existence
Respiratory mechanics have been investigated mathematically for several decades using differential equations models that typically predict air pressure and flow in and between compartments representing aggregate features of the respiratory system. Models have grown in complexity from early compartmental models of dynamic volumes and pressures in the airways, lungs, chest wall, and intrapleural space~\cite{Golden73}. Successive models have built upon this foundation by including nonlinear resistances and compliances, viscoelastic components, and pulmonary circulation~\cite{Verbraak91,Liu98,Athan00}, and more recently been adapted to newborn animal physiology~\cite{LeRolle13}. We previously developed a dynamic nonlinear computational model of infant respiratory mechanics parameterized for the extremely preterm human infant~\cite{Ellwein18} to propose a mechanism of delayed progressive lung volume loss attributed to high chest wall compliance (floppiness)~\cite{Love53,Beltrand08,Kovacs15}.  Our model was the first known attempt to represent these dynamics in premature infants, and also depict the mitigating effects of expiratory laryngeal braking (grunting) and continuous positive airway pressure (CPAP) under simulated high and low chest wall compliance conditions. However, the parameter space contributing overall to breathing dynamics was not explored. Investigation of this fragile demographic is hampered by the challenges associated with clinical data acquisition in a stabilized but extremely preterm infant. Given that ventilation assistance continues to fail in this population and with unknown etiology, this remains an area of continued study~\cite{Manley13,Bhandari13,Siew15}. 

% Need for sensitivity analysis
Forward model simulations using parameter values obtained from experiments or population-based averages provide insight into overall dynamics of a group, but estimating patient-specific parameters requires an optimization algorithm to solve an inverse problem with experimental data. In nonlinear physiological systems with large parameter spaces and a scarcity of data, the optimization problem comes with two inherent challenges for obtaining unique parameter values. The first is parameter {\it sensitivity}, the impact of variation in parameter values on associated model output. A sensitivity analysis can examine the small local changes around each nominal parameter value or the global variability throughout the admissible parameter space~\cite{Eslami94,Karnavas93,LeRolle13,Sher13,Olsen18,Roosa19}. Parameter values may be only valid in a local region, and may affect sensitivities of other parameters in a nonlinear system. The second challenge is parameter {\it identifiability}, either due to the structure of the model (structural) or the availability of data (practical)~\cite{Sher13,Olsen18,Kao18,Roosa19}. In an example of structural unidentifiability, two or more parameters may impact an output in combination (e.g. resistors in series), but determining which parameter had the effect based on the change in output is not possible. Questions regarding parameter identifiability in a pulmonary mechanics model are especially critical in the context of typical clinical data, which for assessment of respiratory mechanics may include only the two dynamic scalar waveforms of volumetric airflow as measured by a pneumotachograph, and pleural pressure as estimated by a pressure transducer in the esophagus. Data acquired under different experimental conditions makes it possible that a waveform from one of the two outputs may change significantly but the other output may show negligible differences, or waveforms may be similar between conditions but mask different underlying dynamics. It is therefore critical to investigate which parameters most influence the model outputs under which conditions, and if any parameter dependencies exist that may allow for simplifying model components. 

% Examples of previous sensitivity analysis and paramater identification
Parameter sensitivity analysis and estimation have been used in previous respiratory and other physiological modeling efforts, with the question of identifiability being explored more recently. However, prior studies focused on simpler linear models in a healthy adult and did not widely address the challenges involved with nonlinear models, non-adult populations, or clinical scenarios. Some of the earliest parameter estimation was performed in several linear respiratory mechanics models~\cite{Lutchen86,Verbraak91,Avanz97,Saatci08}. Schranz et al~\cite{Schranz11,Schranz12a,Schranz12b} applied structural and practical identifiability together with parameter estimation techniques to first-order, viscoelastic, and recruitment models of respiratory mechanics and associated data, but these models represented only a lumped alveolar compartment. In the multi-compartment model of breathing in newborn lambs of Le Rolle et al~\cite{LeRolle13}, the parameter space was explored with Morris screening with respect to flow and pressure outputs including under a continuous positive pressure maneuver. Eight parameters were then estimated per animal using an evolutionary algorithm and continuous time data. A similar approach was used in their follow-up study on an integrated cardiorespiratory model~\cite{AlOmar19} which investigated the effects of CPAP on 
cardiac metrics. While these studies are a few of the only neonatal compartmental models of breathing evident in the literature, it is noted that the model parameters were linear, and the effects of CPAP on respiratory parameter sensitivity and implications of lung health were not considered. 

While the computational model developed by Ellwein Fix et al~\cite{Ellwein18} attempted a nonlinear compliance parameterization and demonstrated respiratory dynamics in premature infants, there was no investigation of sensitivity of the parameter space. Foster et al~\cite{Foster23} applied global screening and local sensitivity analyses to a modified model of thoracoabdominal asynchrony, but that model and analysis were for the different purpose of determining parameters influential on phase angle under a variety of conditions. Similarly, Luke et al~\cite{Luke24} applied parameter inference and model reduction to a simpler compartmental model against a collection of ventilator-induced lung injury pressure-volume curves from rat pups. However, the model did not consider the effects of the chest wall, the data was quasi-static and did not represent spontaneous non-ventilated breathing, and the parameterization for rat pups did not explicitly represent the human preterm infant. Despite recent progress, there remains an opportunity for parameter identification studies specific to nonlinear respiratory mechanics in the preterm infant under clinical conditions.
% 

%final intro
In this study, we propose a reduced respiratory mechanics model based on one previously developed~\cite{Ellwein18} to predict the effect of multiple levels of continuous positive airway pressure (CPAP) on airflow and pleural pressure dynamics of spontaneous breathing in a 1 kg infant. Novelty in this study comes from the analysis of two different scenarios of lung health  common to this demographic, surfactant-treated and surfactant-deficient, defined by distinctly different parameter subsets. To assess global parameter influence and dependencies, we applied Morris screening to parameter ranges that encompass both health scenarios and a clinical range of CPAP. Non-influential components were removed based on Morris results, and the resulting reduced model underwent local derivative-based sensitivity analysis and singular value decomposition (SVD)-based subset selection applied to each health scenario separately over the range of CPAP. Parameter estimation of the resulting sensitive, identifiable parameter set was illustrated with optimizing against clinical waveform data extracted from Abbasi and Bhutani (1990)\cite{Abbasi90}. Optimized parameter values were used to hypothesize about the lung health of the corresponding patient.

%%%%%%%%%% MATHEMATICAL MODELING

\section{Model}\label{sec:Model}

The respiratory mechanics model is represented by lumped-parameter hydraulic compartments describing upper airways ($u$), intermediate collapsible airways ($c$), small peripheral airways ($s$), lungs/alveoli ($A$), intrapleural space ($pl$), and chest wall ($cw$). The lungs are further comprised of elastic ($el$) and a viscoelastic ($ve$) tissue separated in the model by a tissue pressure ($T$). The model is similar to that developed in previous work~\cite{Ellwein18} including the same system of dynamic equations. In this study, the lung recruitment constitutive relation is simplified, and additional constitutive relations are further reduced based on global parameter screening. Following is a description of the state equations, modifications to the constitutive relations, and the nominal parameterization.

%%%%%%%%%% Physiological model description

\begin{figure*}[h]
\centering
\includegraphics[width=\textwidth]{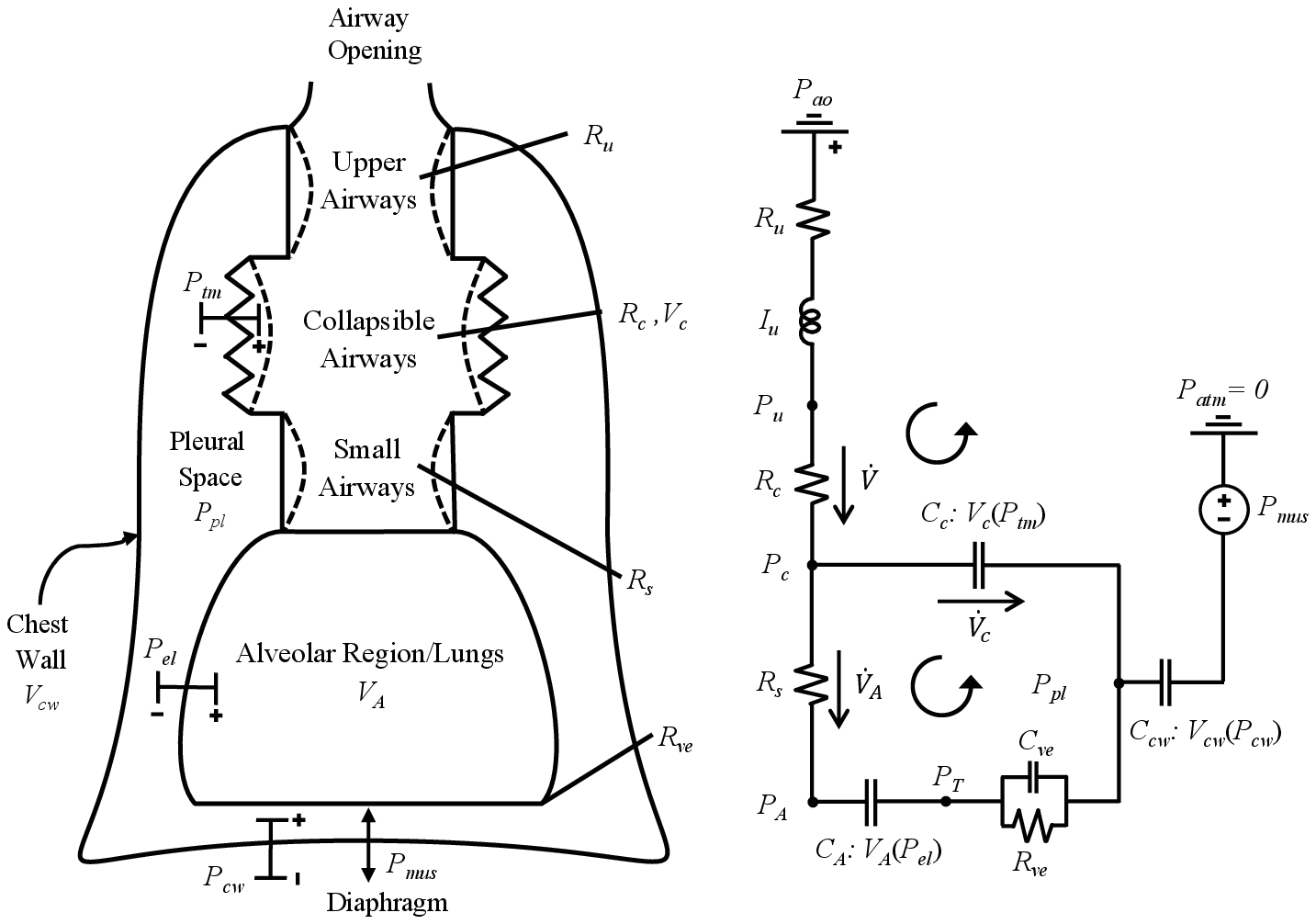}
\caption{Lumped-parameter respiratory mechanics model (left) shown with the electrical analog (right), adapted from Ellwein Fix et al~\cite{Ellwein18}. Each compliant compartment $C$ has an associated volume $V$ as a function of the pressures $P$ across the compartment boundaries. Airflows $\dot{V}$ across resistances $R$ and inertance $I$ are positive in the direction of the arrows. Circular arrows indicate direction of loop summations used to derive the system of differential equations (Eq.~\ref{eq:states}). Subscripts: airway opening $ao$, upper $u$, collapsible $c$, small peripheral $s$, alveolar $A$, viscoelastic $ve$, lung elastic $el$, tissue $T$, transmural $tm$, pleural $pl$, chest wall $cw$, muscle $mus$.}
\label{fig:model}
\end{figure*}

\subsection{State equations}\label{subsec:Model_Equations}

The hydraulic-electric circuit is used to describe dynamic volumes $V(t)$ [ml] and pressures $P(t)$ [cmH$_2$O] within compartments, see Fig.~\ref{fig:model} and Table~\ref{tab:states}. This modeling construct has been used extensively in previous studies in the literature, e.g.~\cite{Golden73,Liu98,Athan00,LeRolle13,Albanese16}. In this style of model, intra-compartment pressures $P_i(t)$ are defined relative to the external atmospheric pressure $P_{ext}=0$, and inter-compartment pressures are defined as the difference between two compartments $P_{ij}=P_i-P_j$ where $j$ is downstream from $i$. Volume rate of change and airflow are represented as $\frac{dV}{dt}$ and $\dot{V}(t)$ [ml/s]  depending on the context. Airflow and pressure gradients are positive in the inward (hydraulic) or counter-clockwise (circuit) direction. \\

% %%% TABLE 1
\begin{table}[!h]
\caption{Descriptions of pressure and volume state variables with units.}
%{\small
\begin{tabular}{lll} 
\hline\noalign{\smallskip}
 	State 		& Units						& Physiologic description					\\ 
 \noalign{\smallskip}\hline\noalign{\smallskip}
  $V$           & [ml]				& Total volume 												\\	
  $V_A$         & [ml]				& Alveolar volume											\\	
  $V_c$         & [ml]				& Collapsible airways volume								\\	
   $V_{cw}$     & [ml]				& Chest wall volume									\\	
 $\dot{V}$      & [ml/s]			& Total Airflow												\\	
$P_u$           & [cmH$_2$O]  & Upper airway pressure \\
$P_c$           & [cmH$_2$O]  & Collapsible airways pressure \\
 $P_{A}$        & [cmH$_2$O]		& Alveolar pressure									    		\\	
$P_T$           & [cmH$_2$O]       & Lung tissue pressure                              \\
 $P_{pl}$       & [cmH$_2$O]		& Pleural pressure											\\	
 $P_{el}$       & [cmH$_2$O]		& Lung elastic recoil		\\	
$P_{ve}$        & [cmH$_2$O]		& Viscoelastic pressure of lung							\\
$P_{l}$     & [cmH$_2$O]	    & Total lung pressure						\\
$P_{tm}$        & [cmH$_2$O]		& Transmural pressure across collapsible airways				\\	
$P_{cw}$        & [cmH$_2$O] 		& Chest wall elastic recoil	pressure							\\	
$P_{mus}$        & [cmH$_2$O]        & Total diaphragm muscle pressure                     \\
\noalign{\smallskip}\hline
\end{tabular}
\label{tab:states}
\end{table}

The compliant compartments, including the collapsible airways, alveoli, and chest wall, have an associated $V_i$ and compliance $C_i$ such that $C_i=\partial V_i / \partial P_{ij}$. Since volumes and pressures are dynamic with respect to time but compliance is not, compliance $C_i$ can be redefined as $\dot{V}_i=C_i\dot{P}_{ij}$. Inter-compartment pressures across compliant boundaries include lung elastic recoil pressure $P_{el}=P_A-P_T$, lung viscoelastic pressure $P_{ve}=P_T-P_{pl}$, transmural airway pressure $P_{tm}=P_c-P_{pl}$, and chest wall recoil pressure $P_{cw}=P_{pl}-P_{mus}$. Kirchhoff's loop rule requires that pressure is conserved around each of the three closed loops of the electrical circuit in Fig. ~\ref{fig:model}, giving the following pressure sums:
\begin{align*}
    0 \quad &= \quad (P_{atm}-P_{ao})+(P_{ao}-P_u)+(P_u-P_c)+(P_c-P_{pl})+(P_{pl}-P_{mus})+P_{mus} \\
    0 \quad &=\quad (P_c-P_A)+(P_A-P_T)+(P_T-P_{pl})+(P{pl}-P_c) \\
    0 \quad & =\quad (P_T-P_{pl})+(P_{pl}-P_T)
\end{align*}
which can be written concisely with inter-compartment pressures as
\begin{align*}
    0\quad &=\quad -P_{ao}+(P_{ao}-P_u)+(P_u-P_c)+P_{tm}+P_{cw}+P_{mus} \\
    0\quad &=\quad (P_c-P_A)+P_{el}+P_{ve}-P_{tm} \\
    0\quad &=\quad (P_T-P_{pl})-P_{ve}
\end{align*}

Pressure drops across resistances $R_i$ [cmH$_2$O$\cdot{}$s/ml] and inertance $I_i$ [cmH$_2$O$\cdot{}s^2$/ml] are defined as typical circuit components with respect to air flows $\dot{V}_i$ and acceleration $\ddot{V}$:
\begin{align*}
    P_{ao}-P_u \quad& = \quad{R_u}\dot{V}+ {I_u}\ddot{V}  \\
    P_u-P_c\quad & = \quad{R_c}\dot{V}  \\
    P_c-P_A\quad& =\quad {R_s}\dot{V}_A \\
    P_T-P_{pl}\quad  &=\quad R_{ve}\dot{V}_{ve} 
\end{align*}
The total system volume $V$ is equivalent to the chest wall volume, modeled as the sum of the alveolar and compressible airway volumes: $V=V_{cw}=V_A+V_c$. Kirchhoff's junction law requires that volume flow is conserved such that $\dot{V}=\dot{V}_A+\dot{V}_c$, where $V_c$ is the volume change of the collapsible airway. \\

Incorporating compliance, resistance, and inertance relationships into the pressure sums gives

\begin{align*}
    0\quad &= \quad -P_{ao}+R_u\dot{V}+I_u\ddot{V}+R_c\dot{V}+P_{tm}+P_{cw}+P_{mus} \\
    0\quad &= \quad R_s\dot{V}_A+P_{el}+P_{ve}-P_{tm} \\
    0\quad &= \quad R_{ve}(\dot{V}_A-\dot{V}_{ve})-P_{ve}
\end{align*}

Reformulating $\dot{V}_{ve}$ in terms of its dynamic compliant state gives the final system of loop equations:

\begin{align*}
    0 \quad &= \quad P_{ao}-R_u\dot{V}-I_u\ddot{V}+R_c\dot{V}+P_{tm}+P_{cw}+P_{mus}\\
    0 \quad &= \quad R_s\dot{V}_A+P_{el}+P_{ve}-P_{tm} \\
    0 \quad &= \quad R_{ve}(\dot{V}_A-C_{ve}\dot{P}_{ve})-P_{ve}
\end{align*}
% The intermediate closed loop serves as a derivation of $\dot{V}_A$ in terms of other system states, mainly consisting of time-varying lung elastic recoil pressure $P_{el}$, viscoelastic lung pressure $P_{ve}$, and the transmural pressure across the compressible airway $P_{ve}$.

Rearranging the loop equations in terms of their derivatives gives the final system of differential equations.
\begin{eqnarray}\label{eq:states}
\ddot{V}&:&\frac{d\dot{V}}{dt}=\frac{1}{I_u}\left(P_{ao}-R_c\dot{V}-P_{tm}-P_{cw}-P_{mus}-R_u\dot{V}\right)\\
\dot{V}_c&:&\frac{dV_c}{dt}=\dot{V}-\dot{V}_A  = \dot{V} - \frac{1}{R_s}\left( P_{tm}-P_{el}-P_{ve}\right) \nonumber\\
\dot{P}_{el}&:&\frac{dP_{el}}{dt}=\frac{\dot{V}_A}{C_A} \nonumber\\
\dot{P}_{ve}&:&\frac{dP_{ve}}{dt}=\frac{\dot{V}_A-(P_{ve}/R_{ve})}{C_{ve}} \nonumber.
\end{eqnarray}

Section \ref{sec:Model}\ref{subsec:Const_relations} introduces nonlinear constitutive relations for states such as $R_c, R_u, P_{tm}, P_{cw}, P_{el}$ and $P_{mus}$ that preserve the well-defined structure of the system.

\subsection{Constitutive relations}\label{subsec:Const_relations}

 Resistances and pressure-volume relations defining compliances assume nonlinear formulations in the initial model, summarized in Table~\ref{tab:functions}. These relations are equivalent to those used in~\cite{Ellwein18}, except for the representation of the recruitment function $F_{rec}$. Previously, in the manner of Hamlington et al~\cite{Hamlington16,Ellwein18}, a fraction of alveoli permanently open at zero pressure and a maximum recruitment fraction were specified. In the current study, we presume that maximum recruitment fraction is 1, and a zero pressure fraction is not prescribed but occurs naturally as a result of mean and range of opening pressures. Parameter descriptions are given in Table~\ref{tab:SSparams}. The quantity $C_A$ used directly in the state equations~\ref{eq:states} is calculated exactly using symbolic computation as $C_A=\p V_A/\p P_{el}$. During spontaneous breathing and no ventilation support, the airway opening pressure $P_{ao}$ is set to 0 cmH$_2$O. The diaphragm muscle pressure $P_{mus}(t)$ connected to the chest wall simulates respiratory muscle activation with a sinusoidal pressure signal.

 % Model architecture in Fig.~\ref{fig:model} is equivalent to that presented in~\cite{Ellwein18}, but with key differences in the nonlinear formulations of certain compartments and eventual model reduction efforts via sensitivity analysis and parameter estimation techniques}. 

%%% TABLE 2
%\begingroup
%\setlength{\tabcolsep}{10pt} % Default value: 6pt
\renewcommand{\arraystretch}{1.5} % Default value: 1
\begin{table}[!h]
\caption{Nonlinear resistance and pressure-volume (compliance) constitutive relations. See Ellwein Fix et al~\cite{Ellwein18} for detailed descriptions.}
\small
\resizebox{\textwidth}{!}{
\begin{tabular}{ll}	
\hline
Description								& Constitutive Relation									\\ 
\hline
Upper airways resistance				& $R_u(\dot{V})=R_{um}+K_u|\dot{V}|$ 								\\
Collapsible airways resistance			& $R_c(V_c)=K_c\left(\ffrac{V_{c,max}}{V_c}\right)^2$			\\
Small (peripheral) airways resistance	& $R_s(V_A)=R_{sd}\cdot e^{K_s(V_A-\text{RV})/(\text{TLC}-\text{RV})}+R_{sm}$,\ \ \ \ $K_s<0$		\\ 
Collapsible airways pressure-volume	& $V_c(P_{tm})=\ffrac{V_{c,max}}{1+e^{-(P_{tm}-c_c)/d_c}}$  	\\
Chest wall pressure-volume			& $V_{cw}(P_{cw})=\text{RV}+b_{cw} \ln \left(1+e^{(P_{cw}-c_{cw})/d_{cw}}\right)$,\ \ \ \ where				\\
                                        & \ \ \ \ \ $V_{cw}(0)=\text{RV}+\nu\cdot \text{VC}$\ \ \ and \ $b_{cw}=\ffrac{\nu\cdot \text{VC}}{\ln(e^{-c_{cw}/d_{cw}}+1)}$.       \\
Lung pressure-volume			&	$V_A(P_{el})=V_{el}(P_{el})\cdot F_{rec}(P_{el})+\text{RV}$,\ \ \ \ where				\\
										& \ \ \ \ \ \ $V_{el}=\text{VC}\cdot(1-e^{(-kP_{el})})$\ \ \ and \ $F_{rec}=\ffrac{1}{1+e^{-(P_{el}-c_F)/d_F}}$. 	\\
Diaphragm muscle driving pressure		& $P_{mus}(t)=A_{mus}\cos (2\pi f t)-A_{mus}$											\\
\noalign{\smallskip}\hline
\end{tabular}}
\label{tab:functions}
\end{table}
%\endgroup

%%% TABLE 3
\begin{sidewaystable}[ph!]
\caption{Parameter descriptions and nominal values for ``healthy'' surfactant-treated lung, a highly compliant rib cage, and no ventilatory support.}
{\footnotesize

 \begin{tabular}{llcll}
\hline\noalign{\smallskip}

Parameter 		& Units    & Baseline Value	& Physiological Description &  References 
\\ 
\noalign{\smallskip}\hline\noalign{\smallskip}
TLC 			& [ml]       &  63		    & Total lung capacity				& \cite{Smith76,Donn98}	\\	
RV 				& [ml]       &  23		    & Residual volume				& \cite{Smith76}	\\	
VC				& [ml]		 &  40        & Vital capacity, TLC-RV			& \cite{Smith76,Donn98}	\\	
RR  			& [br/min]   & 60		& Respiratory rate				    & \cite{Donn98}	\\	
$f$ 			& [br/s]     & 1		& Respiratory frequency, RR/60				& \textemdash	\\	
$T$ 			& [s]        & 1		& Duration of respiratory cycle, $1/f$			& \textemdash	\\	
$A_{mus}$         & [cmH$_2$O] & 1.85      & Breathing muscle pressure amplitude            & \textemdash	\\
$R_{um}$       & [cmH$_2$O/L/s] & 20        & Upper airway resistance minimum value	& \cite{Mortola87,Singh12} \\
$K_u$			& [cmH$_2$O/L$^2$/s$^2$] & 60	& Upper airway resistance flow-dependent coefficient	& \cite{Mortola87,Athan00,Singh12}	\\	
$I_u$ 			& [cmH$_2$O/L/s$^2$] & 0.33	 & Upper airway inertance			& \cite{Singh12,LeRolle13}\\		
$V_{c,max}$ 	& [ml]         & 2.5	    & Collapsible airway maximum volume	& estimated as dead space \cite{Donn98,Neumann15}	\\	
$K_c$ 			& [cmH$_2$O/L/s] & 0.1	    & Collapsible airway resistance volume-dependent coefficient & estimated from adult \cite{Athan00}	\\	
$R_{sm}$ 		& [cmH$_2$O/L/s] & 12		& Small airway resistance minimum		& \cite{Ratjen92,Singh12}	\\	
$R_{sd}$ 		& [cmH$_2$O/L/s] & 20		& Small airway resistance difference		& \cite{Athan00}	\\	
$K_s$ 			& unitless     & -15	    & Small airway resistance volume-dependent coefficient		& \cite{Athan00}	\\	
$c_c$ 			& [cmH$_2$O] & 4.4	    & Collapsible airway pressure at maximum compliance		& \cite{Liu98}	\\	
$d_c$ 			& [cmH$_2$O] & 4.4	    & Collapsible airway pressure range coefficient		& \cite{Liu98}		\\
$k$ 			& [1/cmH$_2$O] & 0.07	& Lung elasticity pressure-dependent coefficient	& \cite{Ferreira11,Hamlington16}  \\
$c_F$			& [cmH$_2$O] &	0.1	    &	Lung recruitment mean opening pressure				& \cite{Hamlington16}  \\
$d_F$			& [cmH$_2$O] &	0.4		&	Lung recruitment pressure range coefficient				& \cite{Hamlington16}  \\
$c_{cw}$ 			& [cmH$_2$O] & 0		& 	Transition pressure, chest wall pressure-volume			& \textemdash	\\	
$d_{cw}$ 			&  [cmH$_2$O] &  0.48	& 	Chest wall pressure-volume slope coefficient 		& 	\textemdash	\\	
$\nu	$		&  [ml]          & 0.25	    & Chest wall relaxation volume coefficient 	& \cite{Donn98,Goldsmith11}	\\	
$C_{ve}$  		& [ml/cmH$_2$O] & 0.005     & Lung viscoelastic compliance	& \cite{Athan00}				\\		
$R_{ve}$  		& [cmH$_2$O/L/s] &20		& Lung viscoelastic resistance		& \cite{Athan00}	\\	
$P_{ao}$        & [cmH$_2$O] & 0         & Airway opening pressure                         & \\

%$V_0$ 			&  35		& $\nu\cdot$VC+RV					& \textemdash	\\	
\noalign{\smallskip}\hline
\end{tabular}}
\label{tab:SSparams}
\end{sidewaystable}

%%%%%%%%%% MODEL PARAMETERS

\subsection{Parameterization}\label{subsec:Model_Parameters}

Nominal parameter values for model equations in Eq.~\ref{eq:states} and Table~\ref{tab:functions}, shown in Table~\ref{tab:SSparams}, comprise the full parameter set $\theta$. Parameter values are the same for surfactant-treated and surfactant-deficient lung scenarios except for the parameter set $\{\text{TLC},k,c_F,d_F\}$. Airway opening pressure $P_{ao}$ is considered a parameter for sensitivity analyses and varies with CPAP level. Further details are given below.

% \subsubsection{Surfactant treated lung ("Treated")}\label{subsubsec:Model_TreatedLung}
\subsubsection{Lung scenarios}\label{subsubsec:Model_TreatedLung}

\paragraph{Surfactant-treated lung ("treated")}

Parameters were manually tuned based on literature reference values such that resulting lung and chest wall compliance curves produced dynamics comparable to key scalar quantities reported for a 1 kg infant with a surfactant-treated lung, a typically underdeveloped and highly compliant (floppy) rib cage, and no ventilatory support~\cite{Ellwein18}. Several key clinical scalar quantities were calculated and compared against literature average values to facilitate parameter tuning.  Table~\ref{tab:vary} reports values for these scalar quantities and references for comparison to literature. Functional residual capacity (FRC) was calculated using a nonlinear solver as the volume where static chest wall pressure $P_{cw}$ balanced against lung elastic pressure $P_{el}$, such that $P_{cw}|_{FRC}+P_{el}|_{FRC}=0$. Tidal volume $V_T$ was calculated as the difference between maximum and minimum lung volume $V_A$ during a single tidal breathing cycle, such that $V_T=\text{max}(V_A)-\text{min}(V_A)$. Minute ventilation, a scalar that describes how much air enters the respiratory system within one minute $\dot{V}_E$\cite{Donn98} was calculated as tidal volume $V_T$ times respiratory rate RR. Breath-to-breath dynamic chest wall compliance $C_{cw,dyn}$ and dynamic lung compliance $C_{l,dyn}$ were calculated during a single breath with the following definitions:
\begin{equation}
    C_{cw,dyn}=\frac{\max(V_{cw})-\min(V_{cw})}{\max(P_{cw})-\min(P_{cw})} \quad \text{and} \quad C_{l,dyn}=\frac{\max(V_{A})-\min(V_A)}{\max(P_{el})-\min(P_{el})}.
\end{equation}
Chest wall compliance is nearly 4 times higher than lung compliance, appropriate for a preterm infant~\cite{Gerhardt80}.

%%% TABLE 4
\begin{table}[ht]
%\begin{adjustwidth}{-1.4in}{0in} % Comment out/remove adjustwidth environment if table fits in text column.
\caption{Model and literature reported scalar quantities referenced for nominal parameter tuning.} 
{\footnotesize
 \begin{tabular}{lcc} 
\hline\noalign{\smallskip}
\multirow{2}{*}{}	Output	     & Model value &    {References} 			\\ 
\noalign{\smallskip}\hline\noalign{\smallskip}
FRC                              & 24.9 ml  & \cite{Smith76,Donn98,Thomas04}  \\
$V_T$                            & 6 ml& ~\cite{Pandit00,Habib03,Schmalisch05}\\
$\dot{V}_E$                       & 360 ml/min & ~\cite{Donn98} \\
$C_{l,dyn}$					 	  & 2.56	ml/cmH$_2$O		& \cite{Gerhardt80,Mortola87,Pandit00}	\\	
$C_{cw,dyn}$						     & 9.9 ml/cmH$_2$O	& \cite{Gerhardt80,Mortola87}	\\	
\noalign{\smallskip}\hline
\end{tabular}
}
\label{tab:vary}
%\end{adjustwidth}
\end{table}

\paragraph{Surfactant-deficient lung ("deficient")}

% \subsubsection{Surfactant deficient lung ("Deficient")}\label{subsubsec:Model_DeficientLung}

\begin{figure}[h]
    \centering
    \includegraphics[width=\textwidth]{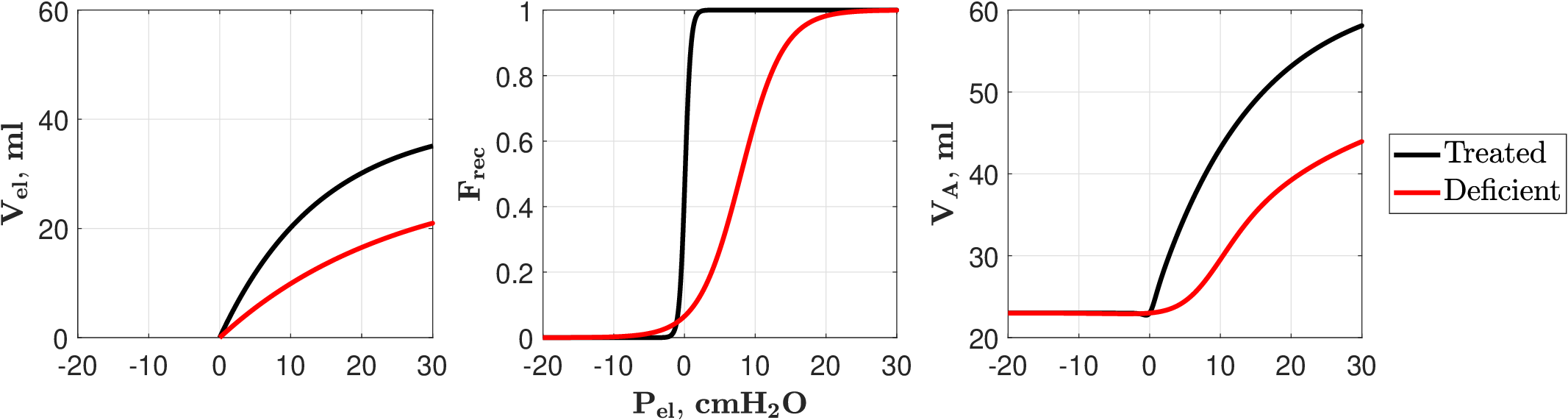}
    \caption{Lung recruitment $F_{rec}(P_{el})$ and pressure-volume relations for $V_{el}(P_{el})$ and $V_A(P_{el})$ with treated (black) and deficient (red) lung modeling scenarios. Lung elastic recoil volume $V_{el}(P_{el})$ (left) represents maximum possible lung volume before scaling by the lung recruitment $F_{rec}(P_{el})$ relation (middle) and summing residual volume RV to obtain alveolar/lung volume $V_A(P_{el})$ (right). See also Table~\ref{tab:functions} for detailed lung pressure-volume relation.}
    \label{fig:lung_ccs}
\end{figure}

The surfactant-deficient lung may have evidence of microatelectasis, or diffuse lung collapse, possibly as a result of insufficient surfactant or other stabilizing forces in the chest wall. In the model, this is described by one or more of the following characteristics: lower TLC, lower elasticity $k$, higher mean opening pressure $c_F$, and larger range of opening pressures $d_F$. These manifest in the pressure-volume curve as lower volumes with respect to alveolar pressure and a right-ward shift of the sigmoidal function $F_{rec}$ in the positive pressure range. In this study, the surfactant deficient lung is modeled by decreasing $k$ and TLC and increasing $c_F$ and $d_F$ (see Table~\ref{tab:lunghealth}). Fig.~\ref{fig:lung_ccs} shows the recruitment relations and static pressure-volume curves for both treated and deficient lungs over a wide range of lung elastic recoil pressures, $P_{el}$. The treated lung recruitment function exhibits a clear distinction between minimum and maximum recruitment at small pressures due to small $c_F$ and $d_F$, leading to higher breath-to-breath dynamic lung compliances $C_{l,dyn}$ at low $P_{el}$ pressures. The deficient lung exhibits reduced lung recruitment over a wide range of $P_{el}$ due to an increase in $c_F$ and $d_F$. When combined with the less compliant $V_{el}(P_{el})$ relation, the static lung pressure-volume relation $V_A(P_{el})$ appears more sigmoidal in construction, inducing much lower $C_{l,dyn}$ at low lung elastic recoil pressures.

%%% Figure 2

%%% TABLE 5
\begin{table}[ht]
%\begin{adjustwidth}{-1.4in}{0in} % Comment out/remove adjustwidth environment if table fits in text column.
\caption{Model parameter values for treated and deficient lungs. See Table~\ref{tab:SSparams} for parameter descriptions. } 
{\footnotesize
 \begin{tabular}{lcc} 
\hline\noalign{\smallskip}
\multirow{2}{*}{}	Parameter	     & Treated &    Deficient 	\\ 
\noalign{\smallskip}\hline\noalign{\smallskip}
TLC                 & 63 ml  & 53 ml \\
$k$                 & 0.07 1/cmH$_2$O& 0.04 1/cmH$_2$O \\
$c_F$               & 0.1 cmH$_2$O & 8 cmH$_2$O\\
$d_F$ 				& 0.5 cmH$_2$O& 3 cmH$_2$O	\\	
\noalign{\smallskip}\hline
\end{tabular}
}
\label{tab:lunghealth}
%\end{adjustwidth}
\end{table}

\subsubsection{Airway opening pressure}\label{subsubsec:Model_Pao}

Continuous positive airway pressure (CPAP) ventilation applies a positive pressure to the patient's airway opening, thereby raising the pressure in the patient's lungs to keep them open and stable. This raises end-expiratory lung volume (EELV) (equivalent to $\min(V_A)$), thereby causing tidal breathing to occur at higher lung pressures. Given the nonlinearity of the constitutive relations, we hypothesized that the application of CPAP would change the relative sensitivities of related parameters and in different ways for the treated vs deficient lung. CPAP was modeled in this study with increasing values of airway opening pressure $P_{ao}$, ranging from 0 to 8 cmH$_2$O~\cite{Locke91,AlOmar19}. Other direct effects of CPAP, such as a lowering of upper airway resistance, were neglected unless they result from constitutive relations.

%%%%%%%%% DATA

\section{Data}\label{sec:data}

Given the fragility of the extremely preterm infant population, it is nearly impossible to obtain clinical waveforms of airflow and pleural pressure from a spontaneously breathing, non-ventilated patient. The majority of these infants require ventilation for breathing stability, and likewise if they are stable enough not to require ventilation, they are not consented for non-critical clinical data acquisition. In a rare study from 1990, Abbasi and Bhutani~\cite{Abbasi90} enrolled 33 healthy, non-ventilated, low birthweight infants (<1500 g and $\leq$34 weeks gestation) to observe and record pulmonary mechanics. The data airflow $\dot{V}^d$ and pleural pressure $P^d_{pl}$ waveforms from a 1 kg infant found in Fig. 1 of ~\cite{Abbasi90} were extracted using PlotDigitizer~\cite{PD}  and used in this study, shown here in Fig. \ref{fig:OptRes}. 

\section{Model Sensitivity and Identifiability}\label{sec:ModelSensitivity_Identifiability}

Sensitivity analyses of the parameter space were performed to show how model outputs respond to physiologically relevant ranges of parameter values under a range of CPAP settings for the model treated and deficient lungs. We first used the Morris global screening method to remove non-influential and non-interacting parameters according to key scalar model outputs. Similar to Colebank and Chesler~\cite{colebank2022silico} and as discussed by Campolongo et al~\cite{Campolongo07}, Morris screening is advantageous for large parameter spaces. Results from the screening were used for model reduction by removing or linearizing compartments. Local derivative-based sensitivity analysis was then applied to the treated and deficient lung reduced parameter spaces according to airflow and esophageal pressure time-dependent outputs, for a range of $P_{ao}$ emulating CPAP. Sensitive parameters were further analyzed for identifiability with an SVD-based subset selection method~\cite{Ipsen11} similar to methods used by Dadashova et al~\cite{Dadashova24}. Model outputs for all aforementioned sensitivity and identifiability analyses were obtained by solving the system in Eq.~\ref{eq:states} with the built-in differential equations solver ode15s in MATLAB 2024b. Absolute and relative tolerances for the solver were fixed at $1\mathrm{e}{-8}$, and the Jacobian finite difference approximation increment used later in subset selection was set to $1\mathrm{e}{-4}$, defined as the square root of the solver tolerance. Time steps were set to 0.002 seconds, and initial conditions for $\dot{V}, V_c,$ and $P_{ve}$ were fixed at 0 ml/s, 0.1 ml, and 0 cmH$_2$O, respectively. The initial condition for $P_{el}$ was set as the lung elastic recoil pressure at functional residual capacity. Further information about the computational model architecture can be found in Appendix A, with full code provided in Data Availability.

\subsection{Morris Screening}\label{subsec:Methods_Morris}

The Morris method \cite{morris1991factorial,saltelli2008global} measures the coarse derivative approximation, or “elementary effect”, of each parameter of interest $\theta_j$ vs a scalar output of interest. The method adjusts one parameter at a time by a scaled step-size $\Delta$, generates a model solution, then calculates an elementary effect, repeated for a prescribed finite number of runs. The mean $\mu_j$ and variance $\sigma_j^2$ of the elementary effects for each parameter is calculated, and the Morris index, defined as $M_j=\sqrt{\mu_j^2+\sigma_j^2}$, ranks parameters by overall influence on scalar model output. Large $\mu_j$ demonstrate that the $j^{th}$ parameter has heavy influence on model output, and large $\sigma_j^2$ demonstrate high nonlinearity or interaction with other parameters\cite{saltelli2008global,qian2020sensitivity}. The computational implementation was based on one developed by Colebank and Chesler~\cite{colebank2022silico} and used in Foster et al~\cite{Foster23}.  Further details regarding the computational scheme can be found in ~\cite{smith2013uncertainty,wentworth2016parameter}. 

Parameter space bounds (Table~\ref{tab:MorrisBounds}) were chosen to be either a feasible physiological range, or $\pm$ 50\% if such a range was unknown. In particular, the upper bounds on $c_F$ and $d_F$ represent a high mean opening pressure and a range of opening pressures such as may occur in a deficient lung. Also $P_{ao}$ extends to a feasible clinical range, to ensure that extremes of CPAP were captured. Morris indices $M_j$ are used to first reduce the model by removal of non-influential parameters, and then set up the parameter set $\theta_{sens}$ on which local sensitivity analysis is applied.

\subsection{Local sensitivity analysis}\label{subsec:Methods_LocalAnalysis}

A single model output is described by the vector:
\begin{equation}
y=[y_p(t_1;\theta),..., y_p(t_{N_i};\theta)]^T
\label{eq:OutputVector}
\end{equation}
where $N_i$ is the number of time points in a single breathing cycle and $\theta$ is the vector of parameters of length $N_j$. Model outputs correspond to data waveforms and thus can be a combination of $y_1=\dot{V}(t)$ and $y_2=P_{pl}(t)$, where $\dot{V}$ is a state variable of Eq.~\ref{eq:states}, and $P_{pl}$ is calculated from state variable $P_{el}$ through several constitutive relations. If both $\dot{V}$ and $P_{pl}$ outputs are analyzed, they are concatenated, giving $y$ length $2N_i$. 
We applied a gradient-based approach~\cite{Eslami94,qian2020sensitivity} to calculate time-dependent sensitivities for each simulation and parameter, around nominal values $\theta^*$ for treated and deficient scenarios. A steady-state full cycle was used to ensure that transient behavior was excluded.

First the output(s) of $y$ are scaled by their maximum absolute value $y_{max}$ (avoiding division by small numbers), 
\begin{eqnarray}
\hat{y}&=&\left[\frac{y(t_1;\theta)}{y_{max}},...,\frac{y(t_{N_i};\theta)}{y_{max}}\right]^T \label{eq:OutputRelative}
\end{eqnarray}

Columns $\hat{S_j}$ of the $N_i \times N_j$ relative sensitivity matrix $\hat{S}$ are determined by differentiating $\hat{y}$ and multiplying by the nominal parameter value $\theta_j$~\cite{Bahill80,Karnavas93,Wu10}: 
\begin{eqnarray}
  \left. \hat{S}_j(t;\theta_j)\right|_{\theta_j=\theta^*_j} &= &
  \left.\theta_j \frac{\p \hat{y}(t_i;\theta_j)}{\p \theta_j} \right|_{\theta_j=\theta_j^*}.
\end{eqnarray} 
Derivatives of $\hat{y}$ with respect to $\theta_j$ were computed with a forward difference approximation using a difference increment of $\epsilon_j=1\mathrm{e}{-4}$~\cite{NCSU,Ipsen11}. To obtain a composite value for ranking, we computed a 2-norm sensitivity $S_j$ across the $pN_i$-size column of $\hat{S}$ for each parameter:
\begin{equation}\label{eqn:sens_normed}
\left. S_{j}= \left\|\hat{S}_j\right\|_2. \right.
\end{equation}

A parameter is considered to be the "sensitive" set for a given lung scenario or CPAP level based on two possible criteria: Quantitatively, the composite sensitivity is greater than the tolerance of the forward difference scheme, or qualitatively, there is an appreciable drop-off in sensitivity when parameters are ranked. The parameter subset on which subset selection is applied is denoted as $\theta_{sub}$.

%%%%%%%%% SUBSET SELECTION

\subsection{Subset selection}\label{subsec:Methods_Subset}

We applied a subset selection method based on singular value decomposition and QR factorization~\cite{Ipsen11,Dadashova24} to further address practical identifiability of the sensitive parameter set given the two clinically relevant outputs~\cite{Pope09}. We began by computing the singular value decomposition of the sensitivity matrix $\hat{S}|_{\theta=\theta^*}=U\Sigma V^T$ where $\Sigma$ is a $N_i\times N_j$ diagonal matrix of singular values in decreasing order, with 0's in rows $N_i-N_j+1$ to $N_i$, and $V$ is a $N_j\times N_j$ matrix containing the corresponding right singular vectors as its columns. The numerical rank $\rho$ indicates the number of maximally independent columns of $\hat{S}$, using a prescribed $\epsilon$ such that $\sigma_{\rho}/\sigma_1\ > \epsilon\geq 10\epsilon_J$, where $\sigma_1$ is the largest singular value and $\epsilon_J$ is the set Jacobian finite difference approximation increment of length $1\mathrm{e}{-4}$ (thus giving a cutoff as $1\mathrm{e}{-3}$). The numerical rank is equivalent to the number of parameters that can be identified given the model output $\hat{y}$ and is used to partition $V=[V_{\rho}V_{N_j-\rho}]$. The particular parameters associated with the $\rho$ largest singular values are found using QR-decomposition with column pivoting. The permutation matrix $P$ that results from the decomposition $V_{\rho}^TP=QR$ is applied to reorder the parameter vector $\theta^*$ to obtain $\tilde{\theta}^*=P^T\theta^*$, 
which is partitioned as $\tilde{\theta}^*=\tilde{\theta}^*_{\rho}\tilde{\theta}^*_{N_j-\rho}$. The vector $\theta_{opt}=\tilde{\theta}^*_{\rho}$ then constitutes an independent set of model parameters deemed estimable as part of a reduced-order optimization problem, introduced in Section~\ref{sec:ModelSensitivity_Identifiability}\ref{subsec:Methods_Opt}.

%%%%%%%%% OPTIMIZATION

\subsection{Optimization}\label{subsec:Methods_Opt}
Patient-specific parameter optimization for the identifiable subset $\theta_{opt}$ was demonstrated by solving the following reduced-order problem by using data for one spontaneously breathing 1 kg infant (see Section~\ref{sec:data}):

\begin{equation}
\arg \min \limits_{\theta} J(\theta_{opt}),
\end{equation}

where $J$ is a cost function, and parameters not found in $\theta_{opt}$ remain fixed at baseline estimates. Nominal parameter values $\theta^*$ for the treated lung were used as initial guesses. Optimization was repeated with deficient lung parameter values to determine if different starting guesses would result in a different optimized parameter set. Parameters not in $\theta_{opt}$ were kept constant at nominal values. Given that, clinically, the patients from that study had no ventilation support, $P_{ao}$ was set at 0 cmH$_2$O.

The optimization included both model outputs and data time series. The resulting cost function $J$ is a least squares formulation using the outputs given in Section~\ref{sec:ModelSensitivity_Identifiability}\ref{subsec:Methods_LocalAnalysis}:
\begin{eqnarray}
  J = \sum_{i=1}^{N} \left|\frac{\dot{V}^m(t_i:\theta)-\dot{V}^d_i}{\dot{V}_{max}^d}\right|^2 + \sum_{i=1}^{N} \left|\frac{P_{pl}^m(t_i;\theta)-P_{pl,i}^d}{P_{pl,max}^d}\right|^2.
\label{eq:min_prob}
\end{eqnarray}
Superscripts {\it d,m} refer to the data and model respectively, and subscript {\it max} denotes the maximum in absolute value of each data set. A bound-constrained Levenberg-Marquardt (L-M) optimization algorithm with trust region~\cite{Ipsen11,Kelley99} and regularization parameter $\gamma=0.2$ was used. Lower and upper bounds for parameter constraints in L-M were set at the Morris bounds, as these were chosen with the specific physiology and feasible clinical conditions in mind. Each parameter value is scaled during the optimization by the difference of the bounds for that parameter. The L-M algorithm terminates with tolerance of $1\mathrm{e}{-4}$ based on gradient norm $||\nabla J(\theta)||$ or residual $J(\theta)$, i.e. the iteration continues until one has fallen below the tolerance for any of the convergence criteria.

\subsection{Summary of Procedures}\label{subsec:Methods_Procedure}

The Morris method was applied with four scalar outputs, chosen to correspond to obtainable clinical measurements:
\begin{itemize}
    \setlength\itemsep{0em}
    \item Tidal volume $V_T$
    \item End-expiratory lung volume $EELV$ $\min(V_A)$
    \item Maximum airflow $\dot{V}_{max}$
    \item Maximum magnitude pleural pressure $|P_{pl}|_{max}$
\end{itemize}
Local sensitivity analyses and subset selection were performed on model outputs emulating possible available clinical data scenarios:
\begin{itemize}
    \setlength\itemsep{0em}
\item Both airflow and pleural pressure signals: time series $\dot{V}(t)$, $P_{pl}(t)$
\item Airflow signal only: time series $\dot{V}(t)$
\item Pressure signal only: time series $P_{pl}(t)$
\end{itemize}
The sequence of sensitivity analysis steps are as follows:
\begin{enumerate}
    \item Perform Morris screening on $\theta$ for all four scalar model outputs with parameter ranges encompassing both treated and deficient lungs, for 1000 runs total. Rank Morris indices $M_j$ for all scalar outcomes and calculate the mean $\bar{M}_j$ and median $\tilde{M}_j$ for each output. 
    \item At this initial screening we consider three tiers of influence $\theta_1,\theta_2,\theta_3$ based on previous practice~\cite{colebank2022silico}, with $\theta_1$ being most influential. If $M_j>\bar{M}_j$ (mean) for one or more scalar model outputs, parameter $\theta_j$ is part of the most influential tier $\theta_1$ and considered for sensitivity analysis. If $M_j<\tilde{M}_j$ (median) for all scalar model outputs, the parameter $\theta_j$ is considered non-influential in $\theta_3$ and is no longer analyzed. All other parameters in between are put in in $\theta_2$ and analyzed further. 
    \item The initial model is reduced by linearizing nonlinear components based on removal of parameters in $\theta_2\cup\theta_3$ from the model but retaining structure.
    \item Local sensitivity analysis is applied to $\theta_{sens}$, a subset of $\theta_1\cup\theta_2$, with respect to simulated $\dot{V}(t)$ and $P_{pl}(t)$ from the reduced model with 5 levels of CPAP ventilatory support such that $P_{ao}=0,2,4,6,8$ cmH$_2$O. Nominal parameter values are set as given in Table~\ref{tab:SSparams} for the treated lung and Table~\ref{tab:lunghealth} to show the differences for the deficient lung for $k$, \text{TLC}, $c_F$, and $d_F$. 
   \item $S_j$ are calculated per parameter for each of 5 CPAP simulations then averaged to obtain a measure of sensitivity for each parameter. The set of parameters with sensitivities higher than an apparent threshold are identified for the subset selection and referred to as $\theta_{sub}$. $S_j$ are also analyzed vs CPAP level to ascertain the nonlinear sensitivity dependence on $P_{ao}$. 
   \item Subset selection is applied to $\theta_{sub}$ for all outputs to discern any identifiability differences between lung health status, and selected parameters form a new  set $\theta_{opt}$.
   \item Sensitive, identifiable parameters in $\theta_{opt}$ are optimized against the data set extracted from~\cite{Abbasi90}.
\end{enumerate}

%%%%%%%%%% RESULTS

\section{Results}\label{sec:results}

%WHAT ARE TRYING TO CONVEY WITH THIS PART OF THE RESULTS
\subsection{Nominal outputs}\label{subsec:results_nominaloutputs}
\begin{figure}[!h]
    \centering
    \includegraphics[width=\textwidth]{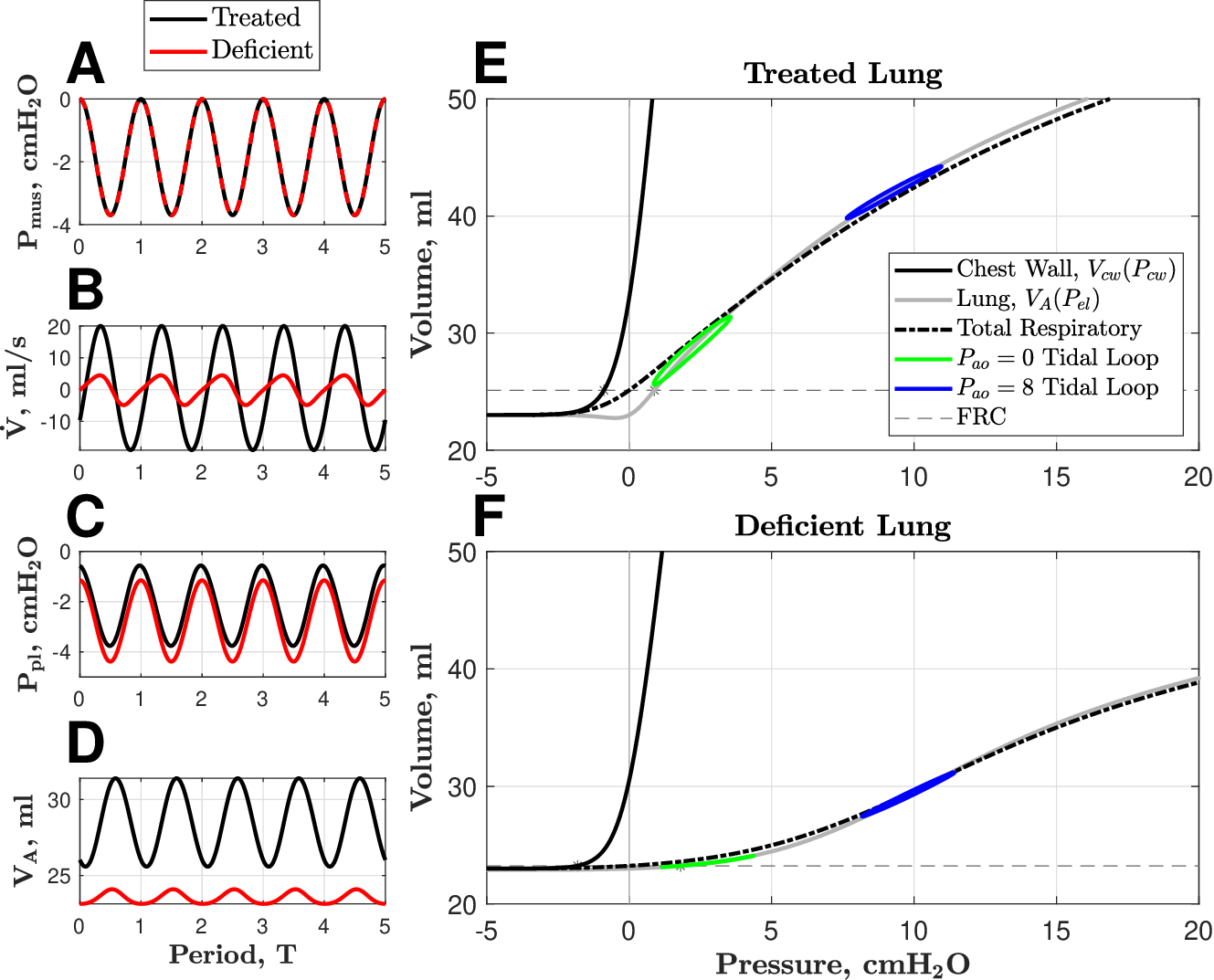}
    \caption{Steady-state waveforms of model outputs with nominal parameter sets for the surfactant-treated and deficient lung scenarios (panels A-D) and static pressure-volume curves for the alveolar/lung region, chest wall, and total respiratory system (panels E-F). Dynamic model outputs include total diaphragm muscle pressure $P_{mus}$, airflow $\dot{V}$, pleural pressure $P_{pl}$, and alveolar/lung volume $V_A$ over five continuous breaths. Chest wall $V_{cw}(P_{cw})$ and lung $V_A(P_{el})$ pressure-volume relations sum pressure-wise to form total respiratory compliance. Tidal loops are represented as alveolar/lung volume $V_A(P_l)$ with fixed $P_{ao}$ values of 0 (green) and 8 (blue) cmH$_2$O, where $P_l$ represents total lung pressure ($P_{el}+P_{ve})$. Functional residual capacity (FRC) is depicted as the volume at which the total respiratory system is at rest, or when respiratory pressure is equivalent to atmospheric pressure.}
    \label{fig:NominalModelOutputs}
\end{figure}

Nominal model outputs for surfactant-treated and deficient lung scenarios with no CPAP ventilation support are shown in Fig.~\ref{fig:NominalModelOutputs}(A-D). Simulated waveforms were generated by the initial model before parameter screening, according to nominal parameter values in Table~\ref{tab:SSparams} and lung health parameters in Table~\ref{tab:lunghealth}. Both surfactant-treated and deficient lung scenarios were driven by the same diaphragmatic muscle pressure signal $P_{mus}$, with the treated lung case reporting a maximum absolute airflow $|\dot{V}|_{max|}$ of 20.03 ml/s, maximum absolute pleural pressure $|P_{pl}|_{max}$ of 3.76 cmH$_2$O, and tidal volume $V_T$ of 5.8 ml. Breath-to-breath dynamic lung and chest wall compliances $C_{l,dyn}$ and $C_{cw,dyn}$ were respectively reported at 2.5 and 10.3 ml/cmH$_2$O, close to reported scalar quantities in Table~\ref{tab:vary} and the 4:1 ratio between $C_{cw,dyn}$ and $C_{l,dyn}$. Compared to the treated lung scenario, the deficient lung model outputs reported severely suppressed airflow and tidal volume, with $|\dot{V}|_{max}$ decreasing to 4.90 ml/s and a near six-fold reduction in $V_T$ to 0.96 ml. Similarly, $C_{l,dyn}$ observed a 10-fold decrease in value to 0.30 ml/cmH$_2$O, and $C_{cw,dyn}$ decreased three-fold to 3.2 ml/cmH$_2$O, effectively stiffening the lung and chest wall compartment boundaries during a single breath. Maximum absolute pleural pressure $|P_{pl}|_{max}$ increased from 3.76 to 4.38 cmH$_2$O, initially suggesting that a stronger inter-compartment pressure exists across the lung boundary. However, this pressure gradient balanced against a stiffer lung due to a decrease in $C_{l,dyn}$, rendering any efforts to improve tidal breathing ineffective.

% associated with the treated lung scenario /  presented in Fig.~\ref{fig:model} according to nonlinear constitutive relations in Table~\ref{tab:functions} and /  whereas those associated with the deficient case were characterized by the key parameter changes outlined in

Static pressure-volume curves for $V_A(P_{el})$ and $V_{cw}(P_{cw})$ with treated and deficient-associated nominal parameters are presented in Fig.~\ref{fig:NominalModelOutputs}(E,F). Total respiratory system compliance is composed of the pressure-wise summation of $V_A(P_{el})$ and $V_{cw}(P_{cw})$, and FRC represents when total respiratory pressure reaches zero.  Tidal breathing loops represent the dynamic trajectory of simulated alveolar volume $V_A$ against total lung pressure $P_l=P_{el}+P_{ve}$. Hysteretic behavior in the tidal loops is caused by the viscoelastic structure of the lungs and their associated parameters $R_{ve}$ and $C_{ve}$. Tidal loops are shown for $P_{ao}$ fixed at 0 and 8 cmH$_2$O, characterizing total breathing dynamics with either absent or extreme CPAP ventilation support, respectively. Increasing $P_{ao}$ from 0 to 8 cmH$_2$O shifts the tidal loop up the static lung pressure-volume curve. FRC was reported to be 25.1 ml in the treated lung scenario and 23.2 ml in the deficient scenario. In the treated lung scenario (E), increasing $P_{ao}$ from 0 to 8 caused a tidal volume decrease from 5.8 to 4.5 ml. Scalar output $C_{l,dyn}$ decreased from 2.5 to 1.5 ml/cmH$_2$O, whereas $C_{cw,dyn}$ increased drastically from 10.3 to 24.0 ml/cmH$_2$O, indicating a stiffer lung and floppier chest wall within the region of tidal breathing. In the deficient lung scenario (F), an increase in $P_{ao}$ caused an improvement in $V_T$ from 0.96 to 3.8 ml and increased $C_{l,dyn}$ and $C_{cw,dyn}$ from 0.30 to 1.24 cmH$_2$O and 3.2 to 11.98 cmH$_2$O, respectively.

%%% FIGURE 3

\subsection{Morris screening}\label{subsec:Results_Morris}

Parameter bounds and Morris screening results are given in Table~\ref{tab:MorrisBounds}. The highest tier for a parameter over all scalar model outputs was designated as its overall tier. The resulting three tiers of parameters were found to be
\begin{equation}
\theta_1=\{P_{ao},k,c_F,d_F,\text{TLC},d_{cw},A_{mus},\text{RV},RR\},
\end{equation}
\begin{equation}
\theta_2=\{R_{um},R_{ve},\nu,R_{sm},V_{c,max}\},
\end{equation}
and 
\begin{equation}
\theta_3=\{c_c,d_c,R_{sd},K_s,K_u,I_u,K_c,C_{ve}\}.
\end{equation}
The set $\theta_1$ was retained in the model as influential, therefore it was not considered for model reduction. The tier with the lowest overall indices $\theta_3$ was deemed non-influential, so parameters in $\theta_3$ were not further analyzed for sensitivity but considered for model reduction. Parameters in the second tier $\theta_2$ are connected to parameters in the first and third tiers through constitutive relations and must be explored further.

\begin{table}[!h]
\caption{Parameter bounds and screening results for Morris sensitivity analysis. 1: Tier one, above the average. 2: Tier 2, below the average but above the median. The highest tier for a parameter over all scalar model outputs was given as its overall tier. Parameters for which the index was never above the median fell into tier 3. Scalar Outputs: $V_T$, tidal volume; EELV, end-expiratory lung volume; $\dot{V}_{max}$, peak airflow; $|P_{pl}|_{max}$, peak magnitude plural pressure. See Table~\ref{tab:SSparams} for parameter descriptions.}
{\small
 \begin{tabular}{lcccccc} 
\hline\noalign{\smallskip}
Parameter   & Bounds	& $V_T$ & EELV & $\dot{V}_{max}$ &  $|P_{pl}|_{max}$   & Morris subset\\ 
\noalign{\smallskip}\hline\noalign{\smallskip}
$P_{ao}$    & [0.01,8]  &  1   &  1    &  1    &  1    & 1 \\
$k$         & [0.04,0.07]&  1  &  1    &  1     &  1   & 1 \\
$c_F$       & [0.1,8]   &  1    & 1   &  1      &  1    & 1 \\
$d_F$       & [0.5,3]   &  1    & 1   &   1     &  1   &  1\\
TLC         & [53,73]   &  1   &  1    &  1    &  2    & 1  \\
$d_{cw}$       & [0.48,2.4] &  1  &  2   &   1     &  1    & 1 \\ %use straight numbers?
$A_{mus}$   & [1.0,4.0] &  1    & 2   &  1      &  1    & 1 \\
RV          & [20,26]   &  2   &  1    &  2    &  2    & 1 \\
RR          & [45,75]   &  2   &  2    &  1    &  2    & 1 \\
$R_{um}$    & [10,30]   &  2    &  2   &   2    &  2    & 2 \\
$R_{ve}$    & [10,30]   &   2   &  2   &  2     &       &  2\\

$\nu$       & [0.2,0.3] &      &  2    &        &  2    & 2 \\ %????
$R_{sm}$    & [6,18]    &  2   &       &   2    &       & 2 \\
$V_{c,max}$ & [2,3]&    &      &       &  2     & 2 \\
$c_c$       & [3.3,5.5] &       &      &       &        & 3 \\
$d_c$       & [3.3,5.5] &       &      &       &        & 3 \\
$R_{sd}$    & [10,30]   &       &       &       &       & 3\\
$K_s$       & [-20,-10] &       &       &       &       & 3\\
$K_u$       & [45,75]   &       &       &       &       & 3\\
$I_u$       & [.23,.43] &       &       &       &       & 3\\
$K_c$       & [.05,.15] &      &       &       &       & 3\\
$C_{ve}$    & [.003,.007]&      &       &       &       & 3\\

\noalign{\smallskip}\hline
\end{tabular}}
\label{tab:MorrisBounds}
\end{table}

\subsection{Model reduction}\label{subsec:Results_ModelReduce}

The initial model was reduced based on removal or consolidation of parameters in the set $\theta_2\cup\theta_3$. After observing all compartments may contribute non-trivially to the system dynamics, the nonlinear portions of several non-influential compartments were linearized instead of removed to maintain model structure. Parameters $R_{um}, K_u, K_c, V_{c,max}, R_{sm}, R_{sd}, K_s$ characterized the upper, collapsible, and small airway resistances and were not contained in $\theta_1$. Recognizing that resistances are essential to the system dynamics, each nonlinear resistance was replaced by a single constant value for $R_u,R_c,R_s$. Parameters $R_{um}, K_u, K_c, R_{sm}, K_s$, and $R_{sd}$ were removed from the remainder of the study. Constant resistances $R_u, R_c,$ and $R_s$ were given nominal values of 20, 0.6, and 12 cmH$_2$O, respectively, informed by average resistance values during tidal breathing. Maximum collapsible airway volume $V_{c,max}$ was retained in the model due to its use in the collapsible airway pressure-volume relation, which was also retained to maintain a volume separate from alveolar volume. Inertance $I_u$ separating resistances $R_u$ and $R_c$ was retained since it has been shown to be a factor in the representation of the very small airways in infant breathing~\cite{Guttmann00, LeRolle13}. After consolidation and removal, parameter sets $\theta_2$ and $\theta_3$ were redefined as
\begin{equation}
    \theta_2=\{R_u,R_{ve},\nu,R_s,V_{c,max},R_c\}
\end{equation}
and
\begin{equation}
    \theta_3=\{c_c,d_c,C_{ve},I_u\}.
\end{equation}

In summary, the reduced model consisted of the same differential equations as the model in Fig.~\ref{fig:model}. The constitutive relations for the airway resistances became constant, and viscoelastic resistance $R_{ve}$ and maximum collapsible airway volume $V_{c,max}$ were retained. For the remainder of this study, we will refer to the "reduced model" as the model presented in Fig.~\ref{fig:model} with nonlinear chest wall, collapsible airway, and lung pressure-volume relations from Table~\ref{tab:functions} and constant resistances $R_u, R_c,$ and $R_s$. Local sensitivity analysis was considered for the set $\theta_1\cup\theta_2$. Parameters in set $\theta_3$ were not included in the local sensitivity analysis.

\subsection{Local sensitivity analysis}\label{subsec:Results_LocalAnalysis}

A subset of $\theta_1\cup\theta_2$,
\begin{equation}
\{RR,P_{ao},\text{RV},R_c\},
\end{equation}
was not included in the sensitivity analysis. Respiratory rate RR would be extracted from data, and $P_{ao}$ would be a model input that mimics CPAP prescribed in the clinic. While RV may change with lung health, it is an extra degree of freedom with TLC and could theoretically be estimated {\it a priori} for the idealized subject. Finally, as $R_c$ and $R_u$ operate in series with a constant inductor, their behavior is dependent, so $R_c$ was fixed at its nominal value since $R_c<<R_u$. The remaining 11 parameters for sensitivity analysis on the reduced model comprise $\theta_{sens}$:
\begin{equation}
\theta_{sens}=\{\text{TLC},c_F,d_F,k,\nu,d_{cw},A_{mus},R_s,R_u,V_{c,max},R_{ve}\}.
\end{equation}

Local sensitivities defined by Eq.~\ref{eqn:sens_normed} were obtained for the parameter set $\theta_{sens}$ against combined and individual model outputs $P_{pl}$ and $\dot{V}$ across five levels of CPAP. All system parameters were initialized by their nominal values in Table~\ref{tab:SSparams}, and specific nominal values for the deficient lung parameters $\text{TLC}, k, c_F$, and $d_F$ are listed in Table~\ref{tab:lunghealth}.

\begin{figure}[!h]
    \centering
    \includegraphics[width=\textwidth]{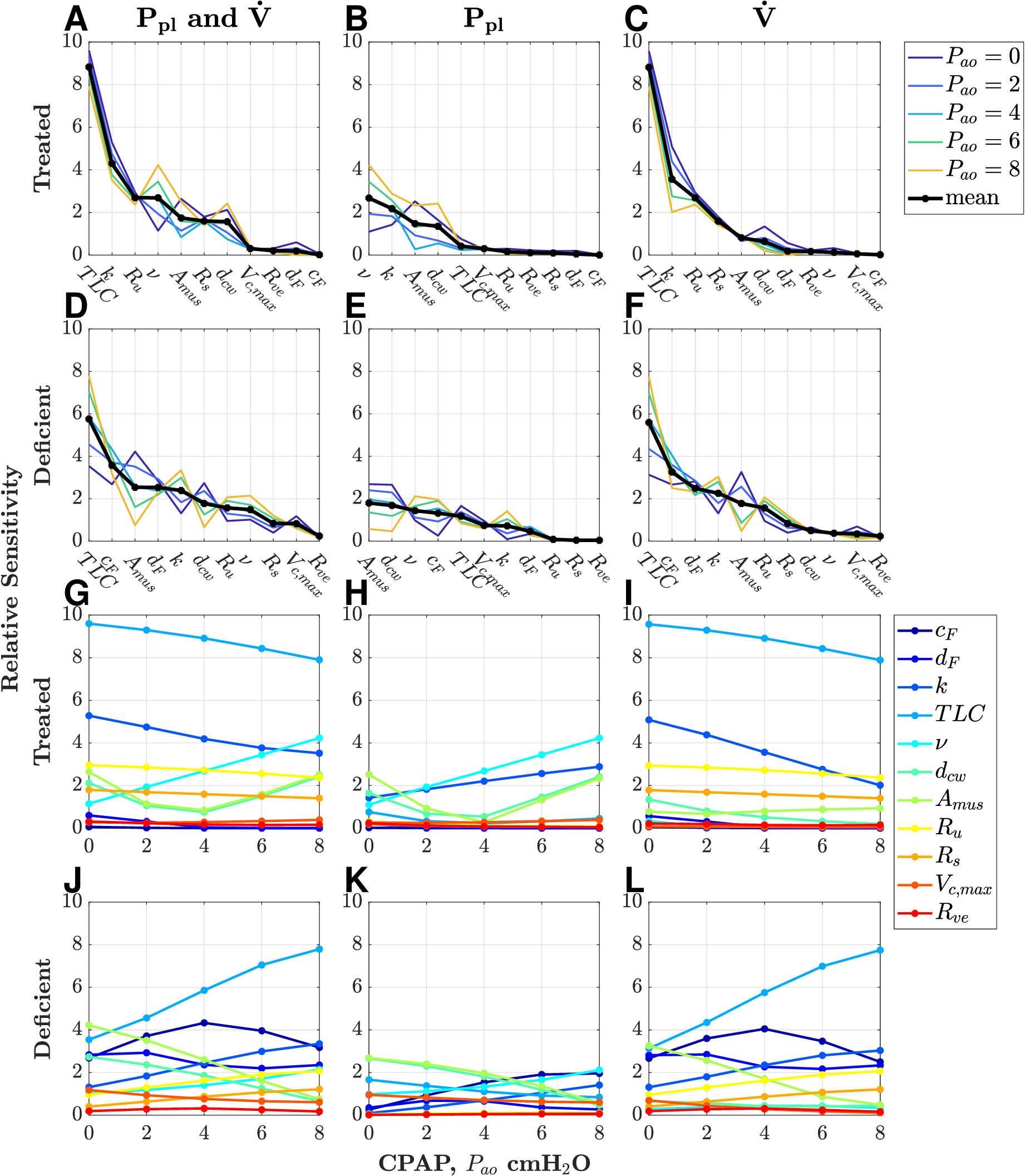}
    \caption{Relative local sensitivities of the parameter set $\theta_{sens}$ shown in decreasing mean sensitivity across 5 levels of CPAP. Simulated scenarios include the surfactant treated (panels A-C) and deficient (panels D-F) lungs with associated nominal parameter values. Panels G-L display relative local sensitivities against CPAP level for each parameter in $\theta_{sens}$. Model outputs of interest during local sensitivity analysis include combined pleural pressure $P_{pl}$ and airflow $\dot{V}$ signals (left), and the individual states $P_{pl}$ (middle) and $\dot{V}$ (right). Administered CPAP is simulated via parameter $P_{ao}$, which is measured in cmH$_2$O units.}
    \label{fig:Sens_Results}
\end{figure}

Figure~\ref{fig:Sens_Results}(A-C) shows the relative local sensitivities of $\theta_{sens}$ for each model output in the treated lung scenario. Total lung capacity TLC, lung elasticity $k$, upper airway resistance $R_u$, chest wall relaxation volume $\nu$, driving pressure amplitude $A_{mus}$, small airway resistance $R_s$, and chest wall compliance coefficient $d_{cw}$ were the parameters with the highest sensitivity rankings against combined airflow $\dot{V}$ and pleural pressure $P_{pl}$ in the treated scenario. A noticeable drop-off in sensitivity occurred where maximum compressible airway volume $V_{c,max}$, viscoelastic resistance $R_{ve}$, lung recruitment range coefficient $d_F$, and mean opening lung pressure $c_F$ had sensitivities less than 10\% of the maximally sensitive TLC. Chest wall relaxation volume, $\nu$, was the most sensitive parameter against pleural pressure $P_{pl}$, but experienced a steep decrease in sensitivity against airflow $\dot{V}$, whereas lung elasticity $k$ was largely sensitive for all individual model outputs.  Figure~\ref{fig:Sens_Results}(D-F) shows relative sensitivities for the parameter set $\theta_{sens}$ for three model outputs in the surfactant-deficient lung scenario. Like in the treated lung scenario, total lung capacity TLC, muscle pressure amplitude $A_{mus}$, lung elasticity $k$, and chest wall compliance slope $d_{cw}$ were found to be the most sensitive parameters against combined airflow $\dot{V}$ and pleural pressure $P_{pl}$ outputs. Lung compliance parameters $c_F$ and $d_F$ had much higher sensitivities and sensitivity rankings in the deficient lung scenario, possibly due to increased nominal values and a flat sigmoidal lung pressure-volume curve. Resistances $R_u,R_s$, and $R_{ve}$ appeared to influence $P_{pl}$ at negligible levels, and $R_{ve}$ was least sensitive against individual and combined model outputs.

Figure~\ref{fig:Sens_Results}(G-L) displays the relative local sensitivities of $\theta_{sens}$ against varying levels of CPAP for the treated and deficient lung modeling scenarios. Relative sensitivities for $\nu$ and $k$ were directly related to administered CPAP level against pleural pressure $P_{pl}$, whereas parameters TLC and $k$ are inversely associated with CPAP level against airflow $\dot{V}$. Furthermore, parameters $A_{mus}$ and $d_{cw}$ displayed an decrease and subsequent increase in relative sensitivity against $P_{pl}$ as $P_{ao}$ increased, indicating preferential influence on model output at either absent or extreme levels of CPAP administration. Parameters $V_{c,max}$ and $R_{ve}$ showed negligible sensitivity against all model outputs and CPAP levels in the treated lung scenario. Resistances $R_u$ and $R_s$ were insensitive towards $P_{pl}$ but highly sensitive towards $\dot{V}$, with little change in contribution as $P_{ao}$ increased. 
When CPAP was not being administered, total lung capacity, lung elasticity, and upper airway resistance were the most sensitive factors in determining pleural pressure and airflow outputs in a treated lung scenario. Parameter impact on the combined model output was generally dominated by impact on airflow. In the deficient lung scenario (panels J-L), sensitivities for TLC and $k$ became directly related to CPAP level $P_{ao}$ against combined outputs, contrasted with the treated lung. Mean opening lung pressure $c_F$ became the second most sensitive parameter at intermediate values of CPAP against combined model outputs, largely determined by its impact on airflow. Muscle pressure amplitude $A_{mus}$ and chest wall pressure-volume slope $d_{cw}$ both showed a significant drop in sensitivity against combined outputs as CPAP increased in the deficient lung scenario, suggesting that chest wall-associated inputs matter less in determining outputs when CPAP is administered.

Table~\ref{tab:SensParams} shows parameter rankings of $\theta_{sens}$ according to mean relative sensitivity index against individual and combined model outputs $P_{pl}$ and $\dot{V}$, corresponding to the mean (black) line in Fig.~\ref{fig:Sens_Results}(A-F). Parameters TLC, $k$, and $A_{mus}$ were, on average, the three most sensitive parameters across all modeling scenarios and outputs, whereas parameters $c_F$ and $d_F$ became markedly more influential towards deficient model outputs. Relative sensitivities and rankings for $V_{c,max}$ were close to the lowest for the combined model output, but it became moderately sensitive against $P_{pl}$. Viscoelastic resistance $R_{ve}$ was consistently the lowest sensitivity parameter for the deficient lung, with slightly more sensitivity in the treated lung. While these results might suggest naming the lowest ranking parameters as insensitive, all sensitivities are higher than the finite difference discretization of $1\mathrm{e}-4$. We theorize that any parameters that might have low enough sensitivities to be undetectable with the finite difference scheme were already screened out by the Morris procedure.  Therefore, all parameters in $\theta_{sens}$ were retained for subset selection. The resulting parameter set used for subset selection, $\theta_{sub}$, is that initially used for local analysis $\theta_{sens}$:
\begin{equation}
\theta_{sub}=\{\text{TLC},k,R_u,\nu,A_{mus},R_s,d_{w},d_F,c_F,V_{c,max},R_{ve}\}.
\end{equation}
These characterize the total lung capacity, lung elasticity, upper airway resistance, chest wall relaxation volume, driving pressure amplitude, small airway resistance, chest wall compliance, lung recruitment range, mean opening lung pressure, collapsible airway volume, and lung viscoelastic resistance.

%%%%%%%%%%%%%% FIGURE 4

%%%%%%%%% TABLE 7 - SENSITIVITY RANKINGS
\begin{table}[!ht]
\caption{Relative sensitivity rankings of the $\theta_{sens}$ parameter set for treated and deficient lung scenarios against combined and individual airflow $\dot{V}$ and pleural pressure $P_{pl}$ states.}
\centering
{\small
 \begin{tabular}{|c|cccccccccccc|} 
\hline\noalign{\smallskip}
& & TLC & $k$ & $R_u$ & $\nu$ & $A_{mus}$ & $R_s$ & $d_{cw}$ & $V_{c,max}$ & $R_{ve}$ & $d_F$ & $c_F$ \\
\noalign{\smallskip}\hline\noalign{\smallskip}
\parbox[t]{2mm}{\multirow{3}{*}{\rotatebox[origin=c]{90}{\textbf{Treated}}}}& $P_{pl}$ \& $\dot{V}$	&  1	    & 2		& 3         & 4     & 5         & 6         & 7     & 8	 & 9 & 10 & 11\\	
& $P_{pl}$ & 5 & 2 & 7 & 1 & 3 & 9 & 4 & 6 & 8 & 10 & 11 \\
& $\dot{V}$ 	&  1		&   2   &  3 & 9 & 5  & 4 &  6 & 10 & 8 & 7 & 11 \\	

\noalign{\smallskip}\hline\noalign{\smallskip}
\parbox[t]{2mm}{\multirow{3}{*}{\rotatebox[origin=c]{90}{\textbf{Deficient}}}}& $P_{pl}$ \& $\dot{V}$ &  1 & 5		&  7        & 8     & 3         & 9         & 6     & 10	& 11 & 4 & 2\\	
& $P_{pl}$ 	&  5		&   7   & 9         	&   3    &   1       &   10       & 2 & 	6 	 & 11 & 8 & 4\\	
& $\dot{V}$	&  1		&  4    &    6     	& 9     &    5        &    7     & 	8 & 10 & 11 & 3 & 2 \\
\noalign{\smallskip}\hline
\end{tabular}}
\label{tab:SensParams}
\end{table}

\subsection{Identifiability}\label{subsec:Results_Identifiability}

The SVD-based method was applied as described in Section~\ref{sec:ModelSensitivity_Identifiability}\ref{subsec:Methods_Subset} to the individual and combined airflow and pleural pressure outputs, given that both outputs contribute to overall model sensitivity. A subset $\tilde{\theta}_{sub,\rho}^*$ was generated for treated and deficient lung scenarios separately and for each clinical data scenario. The method identified eight parameters as identifiable for the deficient lung, but only seven for the treated lung, with the combined model outputs $P_{pl}$ and $\dot{V}$ generally constructing a viable subset by taking the union of the subsets associated with the individual model outputs (see Table \ref{tab:Subset_Pars}). The four non-identifiable parameters in the treated scenario are mean opening lung pressure $c_F$, small airway resistance $R_s$, upper airway resistance $R_u$ and muscle pressure amplitude $A_{mus}$, whereas non-identifiable parameters for the deficient lung are lung elasticity $k$, upper airway resistance $R_u$, and viscoelastic resistance $R_{ve}$. Resistance $R_u$ is likely non-identifiable due to it appearing in series with, and thus sharing a dependence with, $R_s$. Parameters $k$, $A_{mus}$ in the treated lung and $\nu,d_{cw}$ in the deficient lung displayed subset inclusion solely for either combined or individual model outputs, possibly due to interaction between weaker parameters or each other. Taking a less restrictive approach towards a final identifiable subset $\theta_{opt}$ for optimization, all parameters in $\theta_{sub}$ except $R_u$ are candidates for optimization:

% Since we expect that a typical preterm infant lung might have a non-zero mean opening pressure, a clinical data set may be rich enough to identify all 10 sensitive parameters. The parameter set for optimization is therefore the same as for subset selection:

\begin{equation}
\theta_{opt}=\{\text{TLC},k,\nu,A_{mus},R_s,d_{w},d_F,c_F,V_{c,max},R_{ve}\}.
\end{equation}

\begin{table}[!ht]
\caption{Singular value decomposition based parameter subset selection on $\theta_{sub}$ with respect to $\dot{V}$, $P_{pl}$, and combined model outputs, for surfactant-treated and deficient lung scenarios. \textbf{X} denotes inclusion in the parameter subset.}
\centering
{\small
 \begin{tabular}{|c|cccccccccccc|} 
\hline\noalign{\smallskip}
& & TLC & $k$ & $R_u$ & $\nu$ & $A_{mus}$ & $R_s$ & $d_{cw}$ & $d_F$ & $c_F$ & $V_{c,max}$ & $R_{ve}$ \\
\noalign{\smallskip}\hline\noalign{\smallskip}
\parbox[t]{2mm}{\multirow{3}{*}{\rotatebox[origin=c]{90}{\textbf{Treated}}}}& 

$P_{pl}$ \& $\dot{V}$ 	&  \textbf{X}	    & \textbf{X}		&          & \textbf{X}     &          &          & \textbf{X}     & \textbf{X} &	& \textbf{X} & \textbf{X}  \\

& $P_{pl}$ & \textbf{X} &  &  & \textbf{X} & \textbf{X} & & & \textbf{X} & & & \textbf{X} \\

& $\dot{V}$ & \textbf{X} & & & \textbf{X}& \textbf{X}& &\textbf{X}& \textbf{X} & & \textbf{X} & \\	

\noalign{\smallskip}\hline\noalign{\smallskip}
\parbox[t]{2mm}{\multirow{3}{*}{\rotatebox[origin=c]{90}{\textbf{Deficient}}}}& 

$P_{pl}$ \& $\dot{V}$ 	&  \textbf{X}	    &  	&      & \textbf{X}      & \textbf{X}         & \textbf{X}         & \textbf{X}     & \textbf{X}	& \textbf{X} & \textbf{X} &   \\

& $P_{pl}$ & \textbf{X} &  &  &  & \textbf{X}  &   &  &  \textbf{X} & \textbf{X} & &  \\

& $\dot{V}$ & \textbf{X} &  &  & & \textbf{X}& \textbf{X}&  & \textbf{X} &\textbf{X} & \textbf{X} & \\	
\noalign{\smallskip}\hline
\end{tabular}}
\label{tab:Subset_Pars}
\end{table}

\subsection{Parameter estimation}\label{subsec:Results_ParamEstim}

\begin{table}[!ht]
    \caption{Nominal and optimized parameter values for parameter set $\theta_{opt}$ against airflow and pleural pressure data sets extracted from \cite{Abbasi90}. Nominal starting values for the treated and deficient lung scenarios are given to initialize the L-M optimization procedure outlined in section~\ref{sec:ModelSensitivity_Identifiability}\ref{subsec:Methods_Procedure}. Both starting parameter sets converged to the optimal parameter values in the last row. Units are listed in Table \ref{tab:SSparams}. Bolded, italized values indicate the parameter attained its upper constraints; italicized plain text values indicate it attained its lower constraint.}
    \resizebox{\textwidth}{!}{
    \begin{tabular}{|c|c|c|c|c|c|c|c|c|c|c|c|}
    \hline\noalign{\smallskip}
         & TLC & $k$ & $\nu$ & $A_{mus}$ & $R_s$ & $d_{cw}$ & $d_F$ & $ c_F$ & $V_{c,max}$ & $R_{ve}$& Cost\\
         \noalign{\smallskip}
         \hline\noalign{\smallskip}
         \textbf{Treated} & 63 & 0.07  & 0.25 & 1.85 & 12 & 0.48 & 0.5 & 0.1 & 2.5 & 20& 16.771 \\
         \noalign{\smallskip}
         \hline\noalign{\smallskip}
         \textbf{Deficient} & 53 & 0.04 & 0.25 & 1.85 & 12 & 0.48 & 3 & 8 & 2.5 & 20& 36.48 \\
         \noalign{\smallskip}
         \hline\noalign{\smallskip}
        \textbf{Optimal} & \textbf{73} & 0.044 & \textbf{0.28} & 3.24 & \textbf{18} & 1.43 & \textit{0.5} & 2.28 & \textbf{3.0} & 13.09 & 2.092
    \\

        \noalign{\smallskip}
         \hline\noalign{\smallskip}
    \end{tabular}}
    
    \label{tab:opt_pars}
\end{table}

Parameter set $\theta_{opt}$ was optimized against clinically recorded airflow and pleural pressure signals extracted from Abbasi and Bhutani (1990)~\cite{Abbasi90}, using the constrained Levenberg-Marquardt algorithm described in Section ~\ref{sec:ModelSensitivity_Identifiability}\ref{subsec:Results_Identifiability}. Optimization was performed separately using the treated and deficient lung scenarios (see Tables~\ref{tab:SSparams},\ref{tab:lunghealth}) as two distinct nominal parameter sets, and both initial parameter estimates converged to the same optimal values. Optimization constraints were set according to the Morris bounds in Table~\ref{tab:MorrisBounds}. Optimized parameter values were found to be within reasonable physiological ranges, as discussed when constructing the Morris bounds in Section~\ref{sec:ModelSensitivity_Identifiability}. When optimizing the entire parameter set $\theta_{opt}$, all parameter values except lung elasticity $k$ and viscoelastic resistance $R_{ve}$ increased from their nominal treated lung values. Of special notice are total lung capacity TLC, chest wall relaxation volume $\nu$, small airway resistance $R_s$, lung recruitment range coefficient $d_F$, and maximum collapsible airway volume $V_{c,max}$, all of which attained either their higher or lower constraints set by the Morris bounds. Optimized $\theta_{opt}$ values were found to incorporate attributes of both lung health scenarios, with higher TLC and lower $d_F$ reflecting a more treated lung, and lower $k$ and higher $c_F$ reflecting a more deficient lung. The implications of the constraints on data fitting and potential insights on patient-specific health attributes obtained from optimization results are considered in the Discussion.

\begin{figure}[!h]
    \centering
    \includegraphics[width=\textwidth]{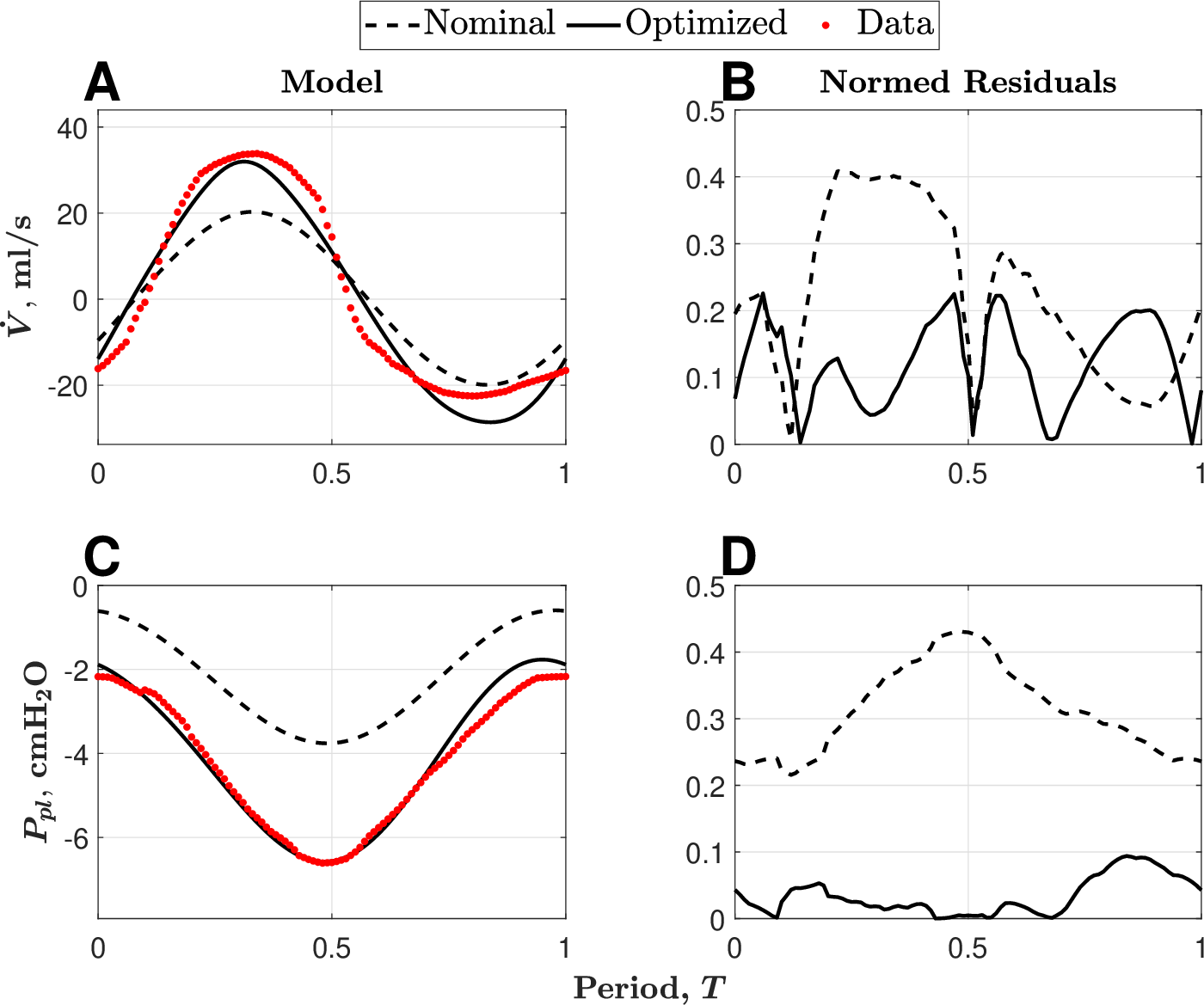}
    \caption{Model outputs generated by nominal treated and optimal parameter sets $\theta$ and $\theta_{opt}$ during spontaneous breathing. A constrained Levenberg-Marquardt optimization scheme with physiologically-informed bounds was applied to the parameters in $\theta_{opt}$ against clinically recorded signals of airflow $\dot{V}$ and pleural pressure $P_{pl}$. Highlighted data in panels \textit{A} and \textit{C} were extracted from waveforms in~\cite{Abbasi90}. Panels \textit{B,D}: Residuals for $\dot{V}$ and $P_{pl}$ were independently normalized by maximum absolute airflow and pleural pressure to maintain optimization on a single scale.}
    \label{fig:OptRes}
\end{figure}

Figure~\ref{fig:OptRes} reports model outputs (A,C) and normalized residual (B,D) plots with nominal and optimized parameter values against clinical data during spontaneous breathing. Model outputs generated by the optimized parameter set $\theta_{opt}$ produced airflow ranging between -28.60 and 31.97 ml/s and pleural pressure between -6.58 and -1.76 cmH$_2$O, both of which represent the ranges in the data set. Airflow and pleural pressure residuals were normalized by their corresponding maximum absolute data values during the optimization scheme to maintain a common scale and discourage the optimizer from favoring one of the model outputs. The normalized residual plot was selected instead of the typical residual to highlight the inner mechanisms of the optimization structure. Normalized residuals were smaller for both optimized model outputs with the exceptions of the 0.7-0.13 and 0.77-0.95 period time frames when the nominal parameter set fits closer to data than the optimal set in the airflow data scenario, likely due to smaller reported airflow during exhalation.

\begin{figure}[!h]
    \centering
    \includegraphics[width=\textwidth]{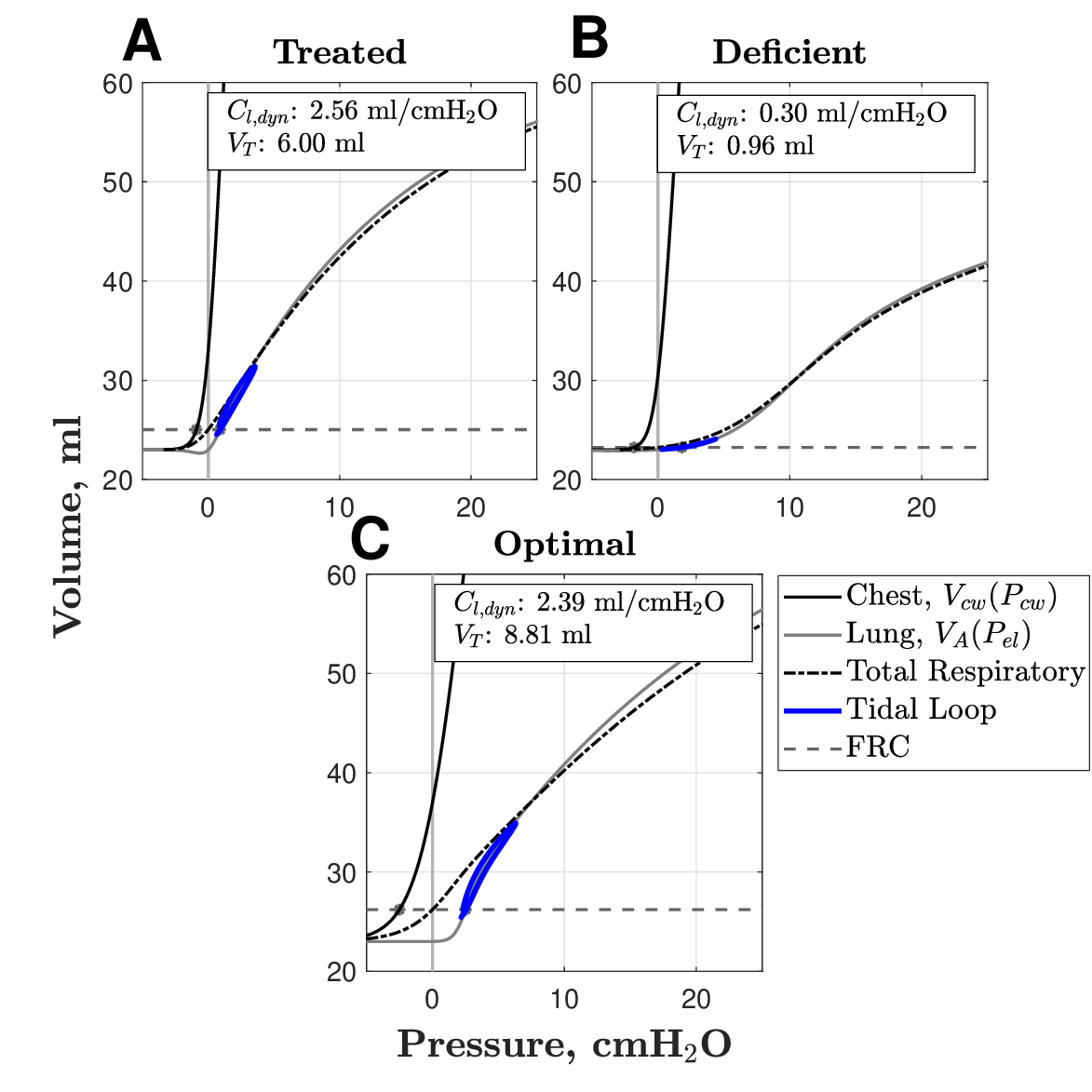}
    \caption{Static pressure-volume relations and tidal breathing loops of the chest wall $V_{cw}(P_{cw})$, lungs $V_A(P_{el})$, and total respiratory system for the reduced model under nominal treated (A), deficient (B), and optimal (C) parameter values. Total respiratory system compliance is generated by the pressure-wise sum of the chest wall and lungs. FRC is determined by the total respiratory system volume when chest wall and lung pressures balance atmospheric pressure, or when the respiratory system pressure is zero. The superimposed blue loop portrays the tidal breathing loop, defined as dynamic $V_A(P_{el}+P_{ve})$. Formulas for dynamic lung compliance $C_{l,dyn}$ and tidal volume $V_T$ are found in Section~\ref{sec:Model}\ref{subsec:Model_Parameters}\ref{subsubsec:Model_TreatedLung}. Optimal parameter values for panel C were obtained in the parameter estimation scheme detailed in Section~\ref{sec:ModelSensitivity_Identifiability}\ref{subsec:Methods_Procedure}}.
    \label{fig:loops}
\end{figure}

Figure~\ref{fig:loops} shows chest wall, lung, and total respiratory pressure-volume relations with treated and deficient lung (see Tables~\ref{tab:SSparams},~\ref{tab:lunghealth}) and optimal parameter scenarios (Table~\ref{tab:opt_pars}). The treated lung scenario in Figure~\ref{fig:loops}A reported a tidal volume of 6.00 ml and respective breath-to-breath dynamic lung $C_{l,dyn}$ and chest wall $C_{cw,dyn}$ compliances of 2.56 ml/cmH$_2$O and 10.09 ml/cmH$_2$O, which match expected values listed in Table~\ref{tab:vary}. Spontaneous inhalation starts on the tidal loop at an a FRC of 25.1 ml at a low elastic lung pressure of 1.046 cmH$_2$O. The deficient lung scenario depicted in Figure~\ref{fig:loops}B reported a severely suppressed tidal volume of 0.96 ml and lower $C_{l,dyn}$ and $C_{cw,dyn}$ values of 0.30 ml/cmH$_2$O and 3.12 ml/cmH$_2$O, respectively. Deficient lung parameters found in Table~\ref{tab:lunghealth} pushed tidal breathing into a lower volume regime, with spontaneous breathing starting close to an FRC of 23.2 ml at a higher lung pressure of 1.807 cmH$_2$O compared to a treated lung. The optimized model scenario shown in Figure~\ref{fig:loops}C reported model outputs similar to those in the treated lung case, with an increased tidal volume of 8.81 ml and dynamic lung compliance $C_{l,dyn}$ of 2.39 ml/cmH$_2$O. The optimal model further reported a dynamic chest wall compliance of 4.74 ml/cmH$_2$O, which is more representative of the deficient lung case. Spontaneous breathing with optimal parameter values began in a higher pressure regime compared to the treated and deficient lung scenarios, with a reported FRC of 26.2 ml at an associated lung pressure of 2.51 cmH$_2$O.

%% FIGURE 5 -- OPTIMIZATION RESULTS

%% FIGURE 6 -- REDUCED MODEL COMPLIANCE CURVES

%%%%%%%%%% DISCUSSION

\section{Discussion}\label{sec:discuss}

This study presented an analysis of the parameter space of a dynamic respiratory mechanics model adapted to the extremely preterm low birthweight infant. In particular, the effects of lung health and simple ventilation support on conventional clinical data tracings were analyzed. Results showed that the combined approach of global screening and local sensitivity and identifiability techniques reduced both the model and the parameter space, to allow prediction of airflow and pleural pressure in a stabilized non-ventilated 1 kg infant. 

%Morris screening advantages and hiccups

The Morris screening was useful for exploring parameter ranges that varied greatly depending on the parameter. As examples, a wide range was set for $P_{ao}$ that encompassed possible clinical settings for CPAP~\cite{Locke91}, and a tight range for $\nu$ knowing where the chest wall pressure-volume curve crosses the volume axis ~\cite{Agostini86}. However, the screening had the freedom to choose parameter combinations that were non-physiological and would not computationally proceed (e.g. an alveolar volume below residual volume would lead to undefined calculations). This was addressed by customizing the Morris code to restart a new search curve when the model would not run. It is possible that the algorithm was inherently less randomized as a result, however given that there were 1000 runs performed, it is more likely that the parameter space was spanned enough for reasonable screening results. Determining the effect of removing problematic runs would be a possible point of future investigation.
 
%Model streamlining and reduction

As respiratory mechanics models have grown in complexity over the decades~\cite{Olender76,Athan00,LeRolle13}, more physiological and pathological conditions have been able to be simulated and investigated computationally. However, the same complexity can obscure which model components are critical to describe the dynamics of interest and potentially add needless computational time. For example, some of the nonlinearity originally in the current model was adapted from a study characterizing the forced vital capacity maneuver which incorporates breathing up to TLC~\cite{Liu98}, but the preterm infant would be breathing only at standard tidal volumes. Considering the purpose of any model allows for only including components and parameters directly related to the outputs under study. The model streamlining and reduction in this study using parameter ranges related to lung health and ventilation allowed for the removal of unnecessary nonlinearity which aided in the parameter identifiability afterwards.

%Sensitivity discussion, treated vs deficient, CPAP dependence

Sensitivity analysis showed that despite normalizing the simulated model outputs, airflow was the greater contributor to sensitivity than the pleural pressure. The parameters characterizing the capacity of the lung, TLC and $k$, were sensitive in both states of lung health. The driving pressure amplitude was a large contributor as well, but more so for pleural pressure than airflow. Additionally, the parameters characterizing the opening ability of the lung were more sensitive when those parameters were higher, signifying a higher pressure and range of pressures needed to open alveoli. The resulting change in the pressure-volume curve of the lung directly impacted the nature of the tidal loop (see Fig.~\ref{fig:loops}). The upper airway resistance had a larger impact in a treated lung, but less so in the deficient lung where it appeared the lung parameters took over. 

The dependence on CPAP levels was highlighted more strongly in the deficient lung, where several parameters such as the resistances had higher sensitivity under no ventilation but then immediately dropped in sensitivity once mild CPAP was applied. Very little of this effect was seen in the pleural pressure, but more strongly in the airflow. Since the deficient lung had much higher opening pressures, it is possible that under no ventilation, the dynamic lung compliance was low (flat pressure-volume curve) which favors the impact of factors external to the lungs, but moving to the "midpoint" of the curve increased dynamic compliance and favored the impact of lung parameters. The decreases of sensitivities of TLC and $k$ with respect to volume in the treated lung but increases in the deficient lung align with this theory. The unusual dip in sensitivity of the muscle driving pressure and chest wall compliance with respect to pleural pressure may also be related to changes in lung compliance along the pressure-volume curve, as this is not seen in the deficient case.

%% Discussion on why certain parameters were removed from sensitvity analysis

Even though non-influential parameters were eliminated early in the parameterization stage of model development, they still provide information about model components. Prime examples include the collapsible airway parameters $c_c,d_c,$ and $K_c$. The physiological analog is likely that preterm infant breathing occurs generally with small tidal volumes and airflow, pushing the typically nonlinear volume-resistance relationship of the collapsible airways into a linear regime. In a global screening method such as the Morris method employed here, linear relationships between parameters would likely reduce their associated elementary effect variances and consequently, their Morris indices. In practice, most other resistance-related parameters such as $R_{sm}, R_{sd}, K_s$, and $K_u$ were globally screened as non-influential likely due to similar reasons. In a small neighborhood of volume and airflow, resistances might be better represented by a linear relationship. Moreover, subset selection results found that constant $R_u$ and $R_s$ exhibited enough dependency to warrant the further use of only one resistance in the parameter optimization process. During local analysis, mean opening lung pressure $c_F$ and lung pressure range $d_F$ were insensitive towards model outputs when given treated lung values, likely since small changes in healthy lung compliance would not drastically change functional residual capacity or tidal breathing characteristics. On the other hand, $c_F$ and $d_F$ values associated with the deficient lung would greatly alter lung pressure at functional residual capacity and dynamic lung compliance in the tidal loop, thus directly affecting the location and magnitude of tidal breathing in pressure-volume space and likely lowering airflow in the upper airways.

%Optimized parameters vs constraints and their meaning

Patient-specific optimization results indicated that it is feasible for this model to identify characterizing parameter values of a non-ventilated 1 kg infant. Tidal volume and dynamic lung compliance of the optimized model are 8.81 ml and 2.39 ml/cmH$_2$O, near the ranges 7.7 to 8.7 ml and 1.99 to 2.13 ml/cmH$_2$O respectively of the extracted data and comparable to values reported in Table~\ref{tab:vary} (see also Fig. 1 of Abbasi and Bhutani)\cite{Abbasi90}. Several parameter values optimized towards or at the treated end of their ranges, indicating a stronger diaphragm, less compliant and more supportive chest wall, higher TLC, and more homogeneous lung. Conversely, the lung elasticity and mean opening pressure parameters optimized away from the treated end of their ranges, suggesting a stiffer lung that requires a higher pressure to open. One hypothesis might be that there is a systemic challenge with opening alveoli across the lung (perhaps an overall surfactant deficiency), as opposed to a larger range of opening pressures indicating varying levels of health (as may occur with a localized injury or pathology.) Future studies should consider stronger approaches for parameter uncertainty and confidence intervals on sparse clinical data, such as parametric bootstrapping intervals~\cite{Banks15, Chowell17}. Estimating model parameters against synthetic data created from simulated model realizations with an added noise structure reflecting measurement error may help address the absence of clinical data altogether~\cite{colebank2022silico,Roosa19,Cho21}.

We noted that some parameters reached their constraints, suggesting that looser constraints might allow for mathematically more accurate parameter estimates. However this must be viewed in light of the model assumptions which could point to further model assessment needed. Regarding lung volume, TLC increases the maximum and $k$ increases the speed at which it reaches the maximum (higher $k$ means higher volumes at lower elastic recoil pressures). In the optimization, TLC reached its upper constraint but $k$ decreases towards its lower constraint, suggesting the maximum lung volume could be larger but it will be reached slowly. A constraint on TLC should be determined in light of possible physiology, e.g. with a nominal value of 63 ml it would not be expected to double in this population. Finally regarding lung pressure, the value for opening pressure range $d_F$ optimized at its lower constraint, suggesting that it might naturally want to be lower. However this would suggest an even tighter range of opening pressures when mean opening pressure is non-negligible at 2.28 cmH$_2$O. It is more likely that a portion of the lung would have higher opening pressures, raising the mean and the range of pressures and reflected by a less steep sigmoidal curve. This constraint was lowered to test the viability of a steeper curve, but caused computational errors from non-physiological scenarios related to the above.

Several features of preterm infant respiratory mechanics are not yet captured by the model and may be contributing to the data fit, particularly the sinusoidal driving pressure function. These include variable frequency of breathing, non-sinusoidal respiratory muscle pressure, intermittent deep breathing (sighing), variable time spent in inspiration vs. expiration, and chemoreflex feedback. Currently the simplest continuous periodic function that does reasonably well emulating breathing in infants is used to drive the model. The data does show a slightly uneven inspiration/expiration ratio, and not purely sinusoidal shape, which could be easily modeled with a different piecewise continuous periodic function. A variety of driving functions have been used to model breathing~\cite{LeRolle13,Albanese16} and could be used here, though they would increase the parameter space for identifiability. Possible parameter dependencies as described above should still be explored further for different model assumptions or physiological scenarios.

%%%%%%% Alternate sensitivity methods

This study focused on global one-at-a-time screening, local differential sensitivity analysis, SVD-based subset selection, and gradient-based optimization for parameter identification. Performing these analyses with two lung health scenarios and five levels of pressure support ventilation explores the issues surrounding the sensitivity of model output to vastly different parameter values in the viable parameter space for the system. There are numerous approaches available to study parameter identification and estimation with optimization that were not used in this study. Many studies approach global analysis as a repeated local analysis using random uniform or Latin hypercube sampling~\cite{Raue13,Olsen18}. The global Morris screening used here is considered an extension of the local analyses and performs more efficiently, making it a reasonable choice for a system with a large parameter space. This could be extended more globally by the calculation of Derivative-based global sensitive measures~\cite{qian2020sensitivity}. Variance-based methods such as the Sobol method and eFAST were not explored, and could add more information, though they are computationally more expensive than the derivative-based methods. Olufsen and Ottesen~\cite{Olufsen13} compare parameter identifiability of a model of heart rate regulation using a structured correlation method, the SVD method, and model Hessian subspace method. Their work found the structured correlation method to produce the ``best'' subset with fewest interdependent parameters, and the SVD method did not give as precise parameter estimates but is much more computationally feasible. Of course the above methods assume deterministic model behavior, but physiological phenomena are stochastic process and may not be as well suited for frequentist techniques. If more probabilistic information about the model is known, methods such as maximum likelihood estimation and Bayesian inference may be considered for future studies~\cite{Kao18,colebank2022silico}.

%Clinical applicability

The clinical applicability of these analyses is directly related to both the available data and the model construction. As seen in Abbasi and Bhutani~\cite{Abbasi90}, lung volume $V_A$ can be a piece of clinically available data, but is recorded as volume relative to FRC instead of an absolute lung volume and in the model is simply integrated airflow. Any breath-to-breath changes in FRC are not captured in this data. Though it was not reasonable to compare the dynamic absolute $V_A$, model tidal volume was used as a scalar output for the Morris screening and could be added to the output vector for parameter estimation against clinical data. It is important to reiterate that we estimated parameters responsible during stable breathing without any ventilation or respiratory distress. Possible effects of other parameters, such as significant airway obstruction represented by high resistances, are not captured by the Morris screening due to the limits on the bounds. A future translational approach could follow what has been done by Kretschmer et al~\cite{Kretschmer17}, who compared model-based parameter estimation with a standard clinical method to determine compliances and resistances under several respiratory maneuvers. From a different perspective, if a parameter determined to be non-identifiable was also deemed clinically relevant, this could motivate collection of new experimental data.

\section{Conclusion}\label{sec:conclusion}
Respiratory mechanics models have been investigated increasingly in recent years. This study investigated the use of practical identifiability and parameter estimation techniques in a lumped-parameter respiratory mechanics model of a spontaneously breathing preterm infant, a demographic that is challenging to study experimentally and has limited data. The model first demonstrates the effects of CPAP administration in surfactant-treated and surfactant-deficient lung scenarios before undergoing global and local sensitivity analyses and model reduction via singular value decomposition and QR-factorization subset selection. A reduced model was achieved primarily by linearizing resistive relations. Optimization was performed against clinical waveforms of airflow and pleural pressure data, resulting in plausible interpretation of individual patient physiology. Results gave a sensitive identifiable parameter subset that may elucidate differences between states of lung health when evaluated using optimization against a clinical data set. These methods may be applied to future data obtained clinically under a variety of states of health and ventilation modalities to estimate patient-specific parameters that may help uncover factors leading to respiratory distress. Future work in the scope of providing parameter uncertainty quantification and responses to adverse clinical scenarios would further inform on possible interventions in a clinical setting. The ability to predict respiratory distress could lead to prevention strategies and assist in the health and stability of the preterm infant population.

\section*{Data Availability}\label{sec:DataAvailability}
Source code that generates model structure and nominal model outputs can be found at \href{http://www.doi.org/10.5281/zenodo.14623545.}{http://www.doi.org/10.5281/zenodo.14623545. } Code pertaining to other processes described in the current study can be made available upon request.

\section*{Conflicts of Interest}
The authors declare no conflicts of interest.

\section*{Acknowledgments}
This research was supported in part by the Atlantic Pediatric Device Consortium FDA grant 5P50FD004193-07.
%\end{acknowledgements}

%\section*{References}

\bibliographystyle{RS}
\bibliography{RSTA}

\section*{Appendix A}

Implementation of the present model in an in-silico environment will be outlined in this section, with corresponding source code provided in the Data Availability section. Section~\ref{sec:ModelSensitivity_Identifiability} briefly described the use of the differential equations solver ode15s to solve the system in Eq.~\ref{eq:states}. However, most variables found in system are not described by their own differential equations, but rather by the nonlinear constitutive relations between them, shown primarily in Table~\ref{tab:functions}. For redundancy, the system in Eq.~\ref{eq:states} defines the change in states such as airflow $\dot{V}$, lung elastic recoil pressure $P_{el}$, collapsible airway volume $V_c$, and viscoelastic lung pressure $P_{ve}$, often as functions of different compartment quantities such as transmural pressure $P_{tm}$, upper airway resistance $R_u$, and others. Algorithm~\ref{Model_Algorithm} outlines the explicit process by which the system in Eq.~\ref{eq:states} was solved in a computational setting, where time \texttt{t}, starting time of the breath cycle \texttt{tprev}, breathing period \texttt{T}, the vector of state solutions at the previous time point \texttt{p}, and the vector of parameter values \texttt{pars} were provided as inputs, and derivative of the states \texttt{dpdt} were outputs. This algorithm was used to construct the main function fed through the ode15s differential equations solver in MATLAB 2024b, as mentioned briefly in Section~\ref{sec:ModelSensitivity_Identifiability}.

\begin{algorithm*}
\caption{\texttt{dpdt = Model(t,p,T,tprev,pars)}}\label{Model_Algorithm}
\begin{algorithmic}[1]

\State \text{\textbf{Extract parameters}}
\Comment{Example: $c_F = \texttt{pars}(1)$, $d_F=\texttt{pars}(2)$}
\State \quad $c_F, d_F, k, \text{TLC}, \text{RV}, RR, \nu, c_{cw}, d_{cw}, A_{mus}$
\State \quad $R_{um}, K_u, K_c, R_{sm}, R_{sd}, K_s, I_u, c_c, d_c, V_{c,max}, C_{ve}, R_{ve}, P_{ao}$
\State{\text{\textbf{Extract state solutions at time} \texttt{t}}}
\State \quad $\dot{V} = \texttt{p}(1)$
\State \quad $P_{el} = \texttt{p}(2)$
\State \quad $V_c = \texttt{p}(3)$
\State \quad $P_{ve} = \texttt{p}(4)$
\State \text{\textbf{Calculate breathing frequency $f$ and $P_{mus}$}}
\State \quad $f=1/\texttt{T}$
\State \quad $P_{mus}=A_{mus}\cos(2\pi f (\texttt{t}-\texttt{tprev}))-A_{mus}$
\Comment{Diaphragm muscle pressure, $P_{mus}$}
\State \textbf{Calculate vital capacity VC and chest wall parameters $a_{cw}, b_{cw}$}
\State \quad VC=TLC-RV
\State \quad $a_{cw}=\text{RV}$
\State \quad $b_{cw}=\frac{\nu \cdot \text{VC}}{\ln(e^{-c_{cw}/d_{cw}}+1)}$
\State \textbf{Calculate alveolar relations}
\State \quad $F_{rec}=\frac{1}{1+e^{-(P_{el}-c_F)/d_F}}$
\Comment{Alveolar recruitment, $F_{rec}$}
\State \quad $V_{el}=\text{VC}(1-e^{-k\cdot P_{el}})$
\Comment{Lung elastic recoil volume, $V_{el}$}
\State \quad $V_A=F_{rec}V_{el}+\text{RV}$
\Comment{Alveolar volume, $V_A$}
\State \quad $C_A=\frac{\text{VC}\cdot k \cdot e^{-k \cdot P_{el}}}{e^{-(P_{el}-c_F)/d_F}+1}-\frac{\text{VC}\cdot e^{-(P_{el}-c_F)/d_F}\cdot (e^{-k \cdot P_{el}}-1)}{d_F\cdot (e^{-(P_{el}-c_F)/dF}+1)^2}$
\Comment{Alveolar compliance, $C_A$}
\State \textbf{Calculate chest wall relations}
\State \quad $V_{cw}=V_A+V_c$
\Comment{Chest wall volume, $V_{cw}$}
\State \quad $P_{cw}=c_{cw}+d_{cw}\cdot \ln(e^{(V_{cw}-a_{cw})/b_{cw}}-1)$
\Comment{Chest wall pressure, $P_{cw}$}
\State \textbf{Calculate pressures}
\State \quad $P_{tm}=c_c-d_c\cdot \ln\left(\frac{V_{c,max}}{V_c}-1\right)$
\Comment{Transmural pressure, $P_{tm}$}
\State \quad $P_l=P_{el}+P_{ve}$
\Comment{Total pulmonary pressure, $P_l$}
\State \quad $P_{pl}=P_{cw}+P_{mus}$
\Comment{Pleural pressure, $P_{pl}$}
\State \quad $P_A=P_l+P_{pl}$
\Comment{Alveolar pressure, $P_A$}
\State \quad $P_c=P_{tm}+P_{pl}$
\Comment{Collapsible airway pressure, $P_c$}
\State \textbf{Calculate resistances}
\State \quad $R_c=K_c\left( \frac{V_{c,max}}{V_c}\right)^2$
\Comment{Collapsible airway resistance, $R_c$}
\State \quad $R_s = R_{sd}\cdot e^{\left(K_s\frac{V_A-\text{RV}}{\text{TLC}-\text{RV}}\right)+R_{sm}}$
\Comment{Small airway resistance, $R_s$}
\State \quad $R_u=R_{um}+K_u|\dot{V}|$
\Comment{Upper airway resistance, $R_u$}
\State \textbf{Calculate remaining relations}
\State \quad $P_u=\dot{V}\cdot R_c+P_c$
\Comment{Upper airway pressure, $P_u$}
\State \quad $\dot{V}_A=(P_c-P_A)/R_s$
\Comment{Alveolar airflow, $\dot{V}_A$}
\State \quad $\dot{V}_c=\dot{V}-\dot{V}_A$
\Comment{'Collapsible airway flow, $\dot{V}$}
\State \textbf{Calculate vector \texttt{dpdt}}
\State \quad $\texttt{dpdt}(1)=\frac{1}{I_u}\left(P_{ao}-P_u-R_u\dot{V}\right)$
\Comment{ODE for $\dot{V},\texttt{p}(1)$}
\State \quad $\texttt{dpdt}(2)=\dot{V}_A/C_A$
\Comment{ODE for $P_{el}, \texttt{p}(2)$}
\State \quad $\texttt{dpdt}(3)=\dot{V}_c$ 
\Comment{ODE for $V_c, \texttt{p}(3)$}
\State \quad $\texttt{dpdt}(4)=(\dot{V}_A-P_{ve}/R_{ve})/C_{ve}$
\Comment{ODE for $P_{ve}, \texttt{p}(4)$}
\end{algorithmic}
\end{algorithm*}

Time-varying solutions to the system in Eq.~\ref{eq:states} were obtained from the outputs of the ode15s solver, upon which dynamic tracings for all other model states could be calculated by rerunning lines 13-37 of Algorithm~\ref{Model_Algorithm} in a separate script over the total time span. Further information on the computational architecture of the current model can be found in the available code.

\end{document}